\documentclass{aastex631}
\usepackage{amsmath}
\usepackage{mathtools}
\usepackage{amssymb}

\shorttitle{Magnetar precession in GRBs}
\shortauthors{Zhang et al.}

\begin{document}
\title{Signature of Triaxially Precessing Magnetars in Gamma-ray Burst X-Ray Afterglows}

\author[0009-0002-0930-1830]{Biao Zhang}
\affiliation{Department of Astronomy, University of Science and Technology of China, Hefei 230026, China; daizg@ustc.edu.cn}
\affiliation{School of Astronomy and Space Science, University of Science and Technology of China, Hefei 230026, China}

\author[0000-0002-1766-6947]{Shu-Qing Zhong}
\affiliation{School of Science, Guangxi University of Science and Technology, Liuzhou 545006, People’s Republic of China}
\affiliation{Department of Astronomy, University of Science and Technology of China, Hefei 230026, China; daizg@ustc.edu.cn}
\affiliation{School of Astronomy and Space Science, University of Science and Technology of China, Hefei 230026, China}

\author[0000-0002-8391-5980]{Long Li}
\affiliation{Department of Astronomy, University of Science and Technology of China, Hefei 230026, China; daizg@ustc.edu.cn}
\affiliation{School of Astronomy and Space Science, University of Science and Technology of China, Hefei 230026, China}

\author[0000-0002-7835-8585]{Zi-Gao Dai}
\affiliation{Department of Astronomy, University of Science and Technology of China, Hefei 230026, China; daizg@ustc.edu.cn}
\affiliation{School of Astronomy and Space Science, University of Science and Technology of China, Hefei 230026, China}

\begin{abstract}
The X-ray afterglows of some gamma-ray bursts (GRBs) exhibit plateaus, which can be explained by the internal dissipation of a newborn millisecond magnetar wind. In the early phase of these newborn magnetars, the magnetic inclination angle undergoes periodic changes due to precession, leading to periodic modulation of the  injection luminosity due to magnetic dipole radiation. This may result in quasi-periodic oscillations (QPOs) on the plateaus. In this paper, we identify four GRBs with regular flux variations on their X-ray afterglow plateaus from Swift/XRT data before November 2023, three of which exhibit periodicity. Based on the likelihood of supporting a precessing magnetar as the central engine, we classify them into three categories: Gold (GRB 060202 and GRB 180620A), Silver (GRB 050730), and Bronze (GRB 210610A). We invoke a model of magnetic dipole radiation emitted by a triaxially freely precessing magnetar whose spin-down is dominated by electromagnetic radiation, to fit the light curves. Our model successfully reproduces the light curves of these four GRBs, including the regular flux variations on the plateaus and their periodicity (if present). Our work provides further evidence for early precession in newborn millisecond magnetars in GRBs.
\end{abstract}


\section{Introduction}
\label{sec:intro}
Gamma-ray bursts (GRBs) are extremely bright and powerful explosions that can originate from the collapse of a massive star (long GRBs) \citep{Woosley1993,MacFadyen&Woosley1999} or the merger of two compact objects (short GRBs), which can be either two neutron stars (NSs) \citep{Paczynski1986,Eichler1989} or a black hole (BH) and a NS \citep{Paczynski1991}. The remnants of GRB explosions, including both long and short bursts, could be rapidly rotating and highly magnetized NSs, referred to as millisecond magnetars, which may serve as central engines for some GRBs \citep{Usov1992,Dai1998PhRvL,Dai1998A&A,Zhang2001,Dai2006,Metzger2011,Bucciantini2012}. A plateau in early afterglow light curves due to energy injection to relativistic forward shocks through magnetic dipole radiation was predicted by \cite{Dai1998PhRvL,Dai1998A&A}, \cite{Zhang2001}, and \cite{Dai2004ApJ}. Subsequently, a significant proportion of early X-ray afterglows from both long and short GRBs detected by Swift/XRT showed a stable and long-duration X-ray plateau, which is considered as evidence for a magnetar providing sustained energy injection as the central engine \citep{Zhang2006,Rowlinson2013,Lv2014,Lv2015}. The X-ray plateaus are followed by a power-law steep decline, and based on different decay indices, they can be classified into \textquotedblleft external plateaus\textquotedblright \,\citep{Lv2015} and \textquotedblleft internal plateaus\textquotedblright \,\citep{Troja2007,Rowlinson2010}. The external plateau is followed by a normal decay phase, which can be explained by an external shock model, where energy is injected by the magnetar through its magnetic dipole radiation \citep{Zhang2006} or by the ejecta with multiple Lorentz factor distributions \citep{Rees&Meszaros1998,Sari&Meszaros2000,Uhm2012}. The most convincing evidence for the magnetar central engine is the existence of internal plateaus, characterized by their most prominent and crucial feature: a sharp decay following the plateau, with time-dependent decay slopes typically steeper than $-3$ and occasionally as low as $-9$ \citep{Troja2007,Liang2007,Lyons2010,Rowlinson2010,Rowlinson2013,Lv2014,Lv2015}. The external shock model is unable to explain such sharp decay, and the sudden drop poses challenges for the black hole central engine model. However, these features of internal plateaus can be well explained by the internal dissipation of magnetar winds, where the sudden drop in luminosity results from the magnetar collapsing into a BH \citep{Troja2007,Liang2007,Rowlinson2010,Rowlinson2013,Lv2014}.

Some X-ray afterglows of GRBs exhibit oscillatory features in their early stages \citep{Dermer1999,Fargion2003,Margutti2008}. If these features are genuine physical phenomena, they could potentially provide further insights into the central engine. The magnetic dipole radiation luminosity of a magnetar depends on the magnetic inclination angle $\alpha$, which is the angle between the magnetic axis and the rotation axis \citep{Spitkovsky2006,KC2009,Philippov2015}. Due to turbulent convection in remnants formed after GRB explosions, which eliminates any pre-existing correlation between the system's magnetic and rotation axes \citep{Thompson&Duncan1993}, magnetars are typically born with a magnetic inclination angle $\alpha$ that is generally not $0\,^\circ$ or $90\,^\circ$ \citep{Melatos2000,Lander&Jones2018}. The evolution of the rotation vector and magnetic field results in the evolution of $\alpha$. In magnetohydrodynamic simulations, $\alpha$ exhibits oscillatory or wobbling behavior \citep{Arzamasskiy2015,Goglichidze2015,Zanazzi2015}, which could potentially lead to oscillatory features observed in the X-ray afterglows of GRBs \citep{Suvorov&Kokkotas2020,ZouLe2021}. The free precession or forced precession of a magnetar can lead to time-dependent variations in $\alpha$, particularly during its early stages, where it may be the primary cause of $\alpha$'s evolution \citep{Goldreich1970,Zanazzi2015}.

The free precession of NSs is supported by observations in some pulsars. The most convincing evidence comes from the isolated pulsar PSR B1828-11, whose pulse shape exhibits long-term, highly periodic variations, with the most prominent period being approximately 500 days \citep{Stairs2000}. These features of PSR B1828-11 may be caused by precession \citep{Jones2001,Link&Epstein2001,Akgun2006}. The observational timing data of PSR B1642–03 also provide support for the free precession of NSs \citep{Cordes1993,Shabanova2001}. The 35-day periodic modulation in the light curve of Her X-1 has also been attributed to precession by some authors \citep{Brecher1972,Kolesnikov2022}. \cite{Zanazzi2015} explained the increase in pulse separation in Crab pulsars \citep{Lyne2013} by the change in magnetic inclination angle caused by pulsar precession. Periodic hard X-ray pulse-phase modulation has been observed in 4U 0142+61, 1E 1547.0-5408, and SGR 1900+14, providing evidence for free precession in magnetars \citep{Makishima2014,Makishima2016,Makishima2021}. Recently, the CHIME/FRB Collaboration discovered a 16.35-day period in FRB 180916 \citep{Chime/FrbCollaboration2020}, which could potentially be attributed to the free precession \citep{Levin2020,Zanazzi&Lai2020} or other types of precession \citep{Sob'yanin2020,Tong2020,Yang&Zou2020,WeiYuJia2022,Wasserman2022} in a magnetar.

Some studies have invoked a biaxially precessing magnetar as the central engine for GRBs and have employed its magnetic dipole radiation to investigate the oscillatory behavior observed in the early X-ray afterglows of GRBs. \cite{Suvorov&Kokkotas2020} used the Akaike Information Criterion (AIC) to demonstrate that for both long GRB 080602 and short GRB 090510, precessing oblique rotating magnetars are more consistent with X-ray afterglow data than non-precessing orthogonal rotating magnetars. Subsequently, \cite{Suvoro&Kokkotas2021} conducted similar analyses on a sample of 25 short GRBs believed to be powered by magnetars. Within their sample, they found that 16 short GRBs strongly supported or favored precessing oblique rotating magnetars as central engines. \cite{ZouLe2021} and \cite{ZouLe2022} invoked a magnetic dipole radiation model of a biaxially precessing magnetar to explain the quasi-periodic oscillations (QPOs) observed in the X-ray afterglows of two long GRBs, GRB 101225A ($4900-7500\,\text{s}$) and GRB 180620A ($200-2300\,\text{s}$). Recently, \cite{ZouLe2024} found a QPO signal with a period of 11 seconds on the internal plateau of the long GRB 210514A. They proposed a possible scenario involving a biaxially precessing supra-massive magnetar model to explain this QPO signal. \cite{ZouLe2022} and \cite{ZouLe2024} considered the QPO signals on the afterglow plateau to be strong evidence of magnetar precession. In our work, we adopt similar criteria to assess the evidence for magnetar precession.

In this paper, our purpose is to search for additional GRBs in the Swift XRT afterglow data that exhibit characteristics similar to those in the GRBs supporting magnetar precession as identified in the works of \cite{ZouLe2022} and \cite{ZouLe2024}. Specifically, we aim to identify GRBs with distinct and regular flux variations on the afterglow plateau, particularly those with QPO signals. Additionally, previous studies on GRBs have typically used a biaxially precessing magnetar as the central engine \citep{Suvorov&Kokkotas2020,Suvoro&Kokkotas2021,ZouLe2021,ZouLe2022,ZouLe2024}, even though magnetar deformations are generally triaxial \citep{Zanazzi2015,GaoYong2020,GaoYong2023}. Therefore, in our work, we adopt a more realistic model of a triaxially precessing magnetar as the central engine for GRBs. We interpret the QPO signals observed on the afterglow plateaus of the GRBs in our study as being caused by variations in the magnetic dipole radiation luminosity due to the quasi-periodic variations in the magnetic inclination angle $\alpha$ during triaxial precession of the magnetar.

The structure of the paper is organized as follows. In Section \ref{sec:radi-prec}, we introduce the magnetic dipole radiation generated by a magnetar undergoing triaxial free precession. In Section \ref{sec:Samp_light}, we present the GRB samples, along with their classification criteria, and provide the fitting results of the precessing magnetar model to the early X-ray afterglow light curves of these GRB samples. Conclusions and discussions are presented in Section \ref{sec:Con}. Throughout the paper, we employ a concordance cosmology characterized by the parameters $H_{0}={\rm 67.4\, km\, s^{-1}\, Mpc^{-1}}$, $\Omega_{\Lambda}=0.68$, and $\Omega_{M}=0.32$ \citep{Planck2018}.

\section{The magnetic dipole radiation of a triaxially precessing magnetar}
\label{sec:radi-prec}
\subsection{Electromagnetic Dipole radiation}
\label{subsec:Electromagnetic}
The energy of GRBs originates from the total rotation energy of newly born millisecond magnetars, which is 
\begin{equation}
E_{\text{rot}}=\dfrac{1}{2}I\Omega^2\simeq2\times10^{52}M_{1.4}R^2_6P^{-2}_{-3}\,{\rm erg},
\label{eq:E_rot}
\end{equation}
where $I\simeq2MR^2/5$ represents the moment of inertia. The rotational angular frequency is given by $\Omega=2\pi/P$,  where \textit{P} denotes the spin period of the magnetar. Additionally, the notations $M_{1.4}=M/1.4\,M_{\odot}$ and $\mathcal{Q}_m=10^{-m}\mathcal{Q}$ in cgs units are adopted throughout the fulltext. 

The electromagnetic radiation and gravitational waves are different ways of consuming the total rotational energy of a magnetar at different rates, both of which can cause the magnetar to spin down \citep{Zhang2001}. If the magnetic dipole spin-down dominates over gravitational wave spin-down, one has \citep{Suvorov&Kokkotas2020},
\begin{eqnarray}
-\dot{E}_{\text{rot}}=-I\Omega\dot{\Omega}=L_{\text{EM}} 
=\dfrac{B^2_{\text{p}}R^6\Omega^4}{6c^3} \lambda(\alpha),
\label{eq:-dotE_rot}
\end{eqnarray}
where $B_{\text{p}}$ is the polar cap surface magnetic field strength. The magnetospheric factor $\lambda$ depends on the magnetospheric environment and the orientation of NS, as determined by the inclination angle $\alpha$ \citep{GJ1969,Spitkovsky2006,KC2009,Philippov2015}. When the magnetospheric environment of NS is the pure vacuum, one has the famous result $\lambda(\alpha)=\sin^{2}\alpha$ \citep{GJ1969}. However, a more realistic scenario may involve the presence of a substantial amount of magnetized plasma within the NS's magnetosphere. Numerical simulations of a magnetosphere filled with charges indicate that $\lambda(\alpha)\simeq1+\sin^{2}\alpha$ \citep{Spitkovsky2006,KC2009,Philippov2015}. Given the constrained comprehension of magnetospheric physics, we adopt a hybrid model \citep{Suvorov&Kokkotas2020,ZouLe2021,ZouLe2022},
\begin{equation}
\lambda(\alpha)\simeq1+k\sin^{2}\alpha,
\label{eq:lambda_alpha}
\end{equation}
where the parameter $k$ quantifies the properties of magnetospheric physics, satisfying $\left|k\right| \leq 1$ \citep{Philippov2014,Arzamasskiy2015}.

The precession of the magnetar results in periodic variations in the magnetic inclination angle $\alpha$, which causes the dipole radiation luminosity to oscillate through the magnetospheric factor $\lambda(\alpha)$, as shown in equations (\ref{eq:-dotE_rot}) and (\ref{eq:lambda_alpha}). After considering precession, beaming effects, and radiative efficiency, the observed isotropic X-ray luminosity is \citep{ZouLe2022}
\begin{equation}
L_{\text{iso,X}}(t)=\dfrac{\eta_{\text{X}}}{f_{\text{b}}}L_{\text{K,0}}\left[1+\dfrac{t}{(1+z)\tau_{\text{sd}}}\right]^{-2}\lambda(\alpha),
\label{eq:L_iso_X_t}
\end{equation}
where $\eta_{\text{X}}$ represents the radiation efficiency within the X-ray band, $f_{\text{b}}=1-\cos\theta_{\text{j}}$ corresponds to the beaming factor, where $\theta_{\text{j}}$ denotes the ejection opening angle, and the decay slope of -2 corresponds to the spin-down of a magnetar through magnetic dipole radiation. Note that $t$ is the time in the observed frame. When the magnetar spin-down is due to magnetic dipole radiation, the initial electromagnetic spin-down luminosity and the characteristic spin-down timescale are
\begin{equation}
L_{\text{K,0}}=1.0\times10^{49}B^{2}_{\text{p,15}}P^{-4}_{0,-3}R^{6}_{6}{\rm \,erg\,s^{-1}}
\label{eq:L_K_0}
\end{equation}
and
\begin{equation}
\tau_{\text{sd}}=2.05\times10^{3}I_{45}B^{-2}_{\text{p,15}}P^{2}_{0,-3}R^{-6}_{6}{\rm \,s},
\label{eq:tau_sd}
\end{equation}
respectively. The spin angular frequency of the magnetar evolves as
\begin{equation}
\Omega=\Omega_{0}\left[1+\dfrac{t}{(1+z)\tau_{\text{sd}}}\right]^{-1/2}.
\label{eq:Omega}
\end{equation}
The obeserved isotropic X-ray Luminosity is related with the observed X-ray flux as
\begin{equation}
L_{\text{iso,X}}=4\pi D^{2}_{\text{L}}F(t)(1+z)^{\Gamma-2},
\label{eq:L_iso_X}
\end{equation}
where $D_{\text{L}}$ is the luminosity distance, $z$ is the redshift, and $\Gamma$ is the X-ray spectral index.

\subsection{Triaxial free precession of an NS}
\label{subsec:precession}
The newly born NSs are typically not perfectly spherical due to the influence of various physical factors, which can cause deviations from a spherical shape \citep{Zanazzi2015,GaoYong2020,GaoYong2021,GaoYong2023}. The elasticity of the crust and the strong internal magnetic fields are the major causes of the deformation of newborn NS \citep{GaoYong2023}. We regard the NS as a rigid body undergoing triaxial deformation. When the principal axis of the NS is misaligned with its rotation axis, the rotating deformed NS will precess around the total angular momentum \citep{landau1960mechanics,GaoYong2020}.

When a deformed, rigid-body NS is not subjected to external torques, it undergoes free precession, with its motion governed by the Euler equations in the body frame, which is \citep{landau1960mechanics}
\begin{equation}
\frac{d\boldsymbol{L}}{dt}+\boldsymbol{\Omega}\times\boldsymbol{L}=0,
\label{eq:dL_dt}
\end{equation}
where $\boldsymbol{\Omega}$ is the angular velocity vector, $\boldsymbol{L}=\boldsymbol{I}\cdot\boldsymbol{\Omega}$ represents the angular momentum with $\boldsymbol{I}$ being the moment of inertia tensor, and the derivative $d/dt$ is taken in the body frame. We use these directions of the principal axes of inertia as the coordinate axes in the corotating body frame. Let $\boldsymbol{\hat{e}_{1}}$, $\boldsymbol{\hat{e}_{2}}$, and $\boldsymbol{\hat{e}_{3}}$ represent the three principal axes (or coordinate axes of the body frame), where $I_{1}<I_{2}<I_{3}$ are their respective principal moments of inertia. The angular momentum and the angular velocity in the body frame are expressed as $\boldsymbol{L}=L_1\boldsymbol{\hat{e}_{1}}+L_2\boldsymbol{\hat{e}_{2}}+L_3\boldsymbol{\hat{e}_{3}}$ and $\boldsymbol{\Omega}=\Omega_1\boldsymbol{\hat{e}_{1}}+\Omega_2\boldsymbol{\hat{e}_{2}}+\Omega_3\boldsymbol{\hat{e}_{3}}$, respectively.

\begin{figure}
\centering
\includegraphics[width=0.6\textwidth, angle=0]{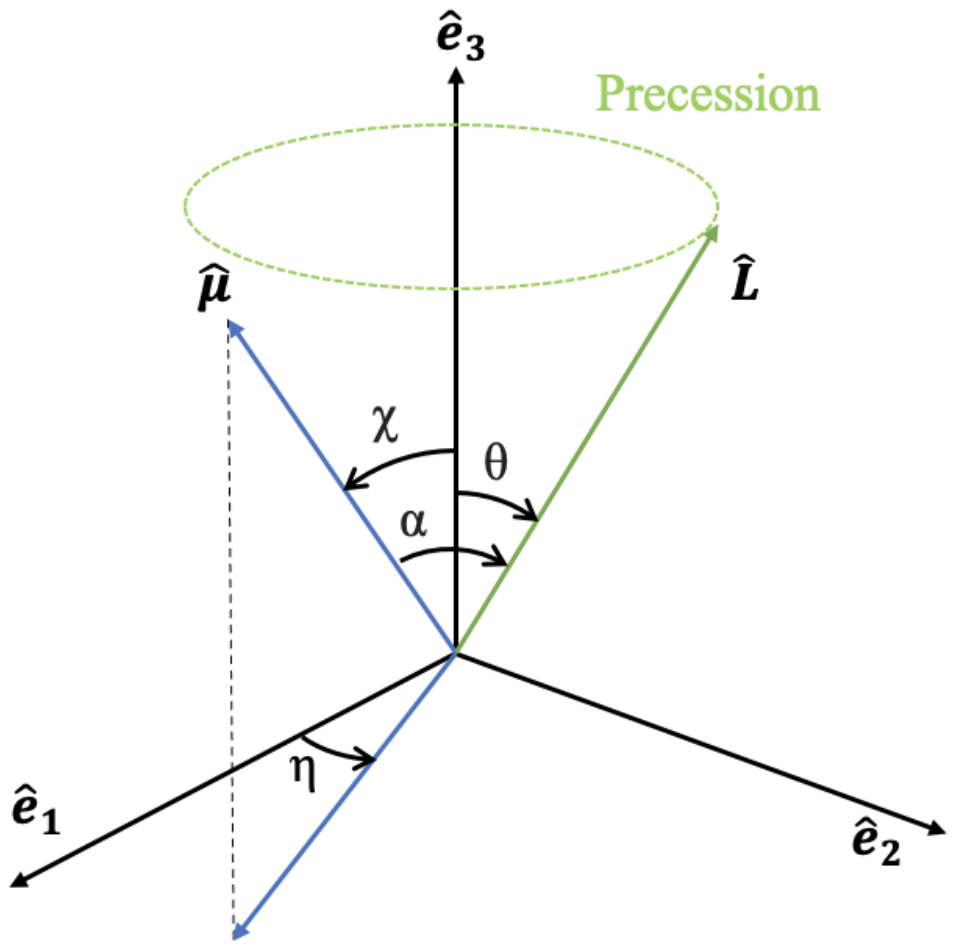}
\caption{The geometry of NS precession in the body frame. The unit angular momentum $\boldsymbol{\hat{L}}$ precesses around $\boldsymbol{\hat{e}_{3}}$, and the angle $\theta$ between them undergoes nutation during precession. The unit dipole moment $\boldsymbol{\hat{\mu}}$ remains constant in the body frame, with its polar angle and azimuthal angle denoted as $\chi$ and $\eta$, respectively. The angle between the unit angular momentum $\boldsymbol{\hat{L}}$ and the unit dipole moment $\boldsymbol{\hat{\mu}}$ is represented by $\alpha$.}
\label{Fig_precession}
\end{figure}

In the case of a freely precessing NS, both its kinetic energy and angular momentum remain conserved, as described by the following equations:
\begin{equation}
2E=I_{1}\Omega^{2}_{1}+I_{2}\Omega^{2}_{2}+I_{3}\Omega^{2}_{3}
\label{eq:2E}
\end{equation}
and
\begin{equation}
L^{2}=I^{2}_{1}\Omega^{2}_{1}+I^{2}_{2}\Omega^{2}_{2}+I^{2}_{3}\Omega^{2}_{3},
\label{eq:L^2}
\end{equation}
where $E$ is the rotational energy, $L$ is the magnitude of the angular momentum. Equation (\ref{eq:dL_dt}) satisfies both of these equations. 

Using Jacobian elliptic functions, one can obtain a well-known and accurate theoretical solution to Equation (\ref{eq:dL_dt}) \citep{landau1960mechanics,Akgun2006,Zanazzi2015,GaoYong2020,Wasserman2022,Kolesnikov2022,GaoYong2023}. At the initial time $t=0$, we adopt the usual settings and set $\Omega_2=0$. In the following, we assume $L^2>2EI_2$, which implies that the angular momentum $\boldsymbol{L}$ precesses around $\boldsymbol{\hat{e}_3}$ in the body frame, as shown in Figure \ref{Fig_precession}. The results of other assumptions can be obtained by correctly reordering the indices \citep{Akgun2006,Zanazzi2015,GaoYong2023}. \cite{landau1960mechanics} offered the theoretical solution to Equation (\ref{eq:dL_dt}), which describes the temporal evolution of angular velocity components in the body frame: 
\begin{equation}
\Omega_{1}=\sqrt{\frac{2EI_3-L^2}{I_1(I_3-I_1)}}\,{\rm cn}(\tau,m),
\label{eq:Omega_1}
\end{equation}
\begin{equation}
\Omega_{2}=\sqrt{\frac{2EI_3-L^2}{I_2(I_3-I_2)}}\,{\rm sn}(\tau,m),
\label{eq:Omega_2}
\end{equation}
\begin{equation}
\Omega_{3}=\sqrt{\frac{L^2-2EI_1}{I_3(I_3-I_1)}}\,{\rm dn}(\tau,m),
\label{eq:Omega_3}
\end{equation}
where dn, sn, and cn are the Jacobian elliptic functions. Here $\tau=\Omega_{\text{P}}t$ represents the dimensionless time, with $\Omega_{\text{P}}$ defined as
\begin{equation}
\Omega_{\text{P}}=\sqrt{\frac{(I_3-I_2)(L^2-2EI_1)}{I_1I_2I_3}}.
\label{eq:Omega_P}
\end{equation}
The Jacobian elliptic functions parameter $m$ is determined by the following expression:
\begin{equation}
m=\frac{(I_2-I_1)(2EI_3-L^2)}{(I_3-I_2)(L^2-2EI_1)}.
\label{eq:m}
\end{equation}
It's important to note that $m<1$, because of the assumption made above that $L^2>2EI_2$. One can also derive the temporal evolution of the angular momentum components in the body frame, from equations (\ref{eq:Omega_1}), (\ref{eq:Omega_2}), and (\ref{eq:Omega_3}), using the relationship $L_{\text{i}}=I_{\text{i}}\Omega_{\text{i}}$, where $\text{i}$ corresponds to subscripts 1, 2, and 3. The angular velocity components and the angular momentum components in the body frame are periodic due to the periodicity of the Jacobian elliptic functions, with a period
\begin{equation}
T=4K(m)\sqrt{\dfrac{I_1 I_2 I_3}{(I_3-I_2)(L^2-2E I_1)}},
\label{eq:T}
\end{equation}
where $K(m)$ represents the first kind complete elliptic integral. The period $T$ represents the period of free precession.
One needs to note that the Jacobian elliptic functions are $4K(m)$ periodic, not $2\pi$ periodic, so the relation $T=4K(m)/\Omega_{P}$ holds.

Following \cite{GaoYong2023} and \cite{Akgun2006}, we simplify the solutions to equation (\ref{eq:dL_dt}) as follows. To better describe the precession motion of the deformed NS, we introduce the following parameters
\begin{equation}
\epsilon_2=\dfrac{I_2-I_1}{I_1},
\label{eq:epsilon_2}
\end{equation}
\begin{equation}
\epsilon_3=\dfrac{I_3-I_1}{I_1},
\label{eq:epsilon_3}
\end{equation}
\begin{equation}
\delta=\dfrac{I_3(I_2-I_1)}{I_1(I_3-I_2)},
\label{eq:delta}
\end{equation}
\begin{equation}
\theta=\arccos\dfrac{L_3}{L}.
\label{eq:theta}
\end{equation}
Here, $\epsilon_2$ and $\epsilon_3$ quantify the degree of non-sphericity, $\delta$ characterizes the degree of triaxiality or the deviation from axisymmetry, and $\theta$ denotes the angle between $\boldsymbol{L}$ and $\boldsymbol{\hat{e}_3}$ known as the wobble angle. In the biaxial scenario, the wobble angle $\theta$ remains unchanged, while in the triaxial case, the wobble angle $\theta$ undergoes nutation with a period of $T/2$. Given that $I_{1}<I_{2}<I_{3}$, it follows that $\epsilon_2$ is smaller than $\epsilon_3$. Using the parameters defined above, we can get
\begin{equation}
I_2=I_1(1+\epsilon_2),
\label{eq:I_2}
\end{equation}
\begin{equation}
I_3=I_1(1+\epsilon_3),
\label{eq:I_3}
\end{equation}
\begin{equation}
\delta=(1+\epsilon_3)\dfrac{\epsilon_2}{\epsilon_3-\epsilon_2}.
\label{eq:delta_2}
\end{equation}
The $\Omega_{\text{P}}$ and the parameter $m$ can be reexpressed as:
\begin{equation}
\Omega_{\text{P}}=\dfrac{\epsilon_{3}L\cos\theta_0}{I_{3}\sqrt{1+\delta}},
\label{eq:Omega_P_2}
\end{equation}
and
\begin{equation}
m=\delta\tan^{2}\theta_0,
\label{eq:m_2}
\end{equation}
where $\theta_0$ is the wobble angle at the initial time $t=0$.

The solution to equation (\ref{eq:dL_dt}) can also be expressed as the temporal evolution of the unit angular momentum vector components in the body frame, as functions of the parameters defined above, as follows:
\begin{equation}
\hat{L}_1=\sin\theta_{0}\,{\rm cn}(\Omega_{{\rm P}}t,m),
\label{eq:hat_L_1}
\end{equation}
\begin{equation}
\hat{L}_2=\sin\theta_{0}\sqrt{1+\delta}\,{\rm sn}(\Omega_{{\rm P}}t,m),
\label{eq:hat_L_2}
\end{equation}
\begin{equation}
\hat{L}_3=\cos{\theta}=\cos\theta_{0}\,{\rm dn}(\Omega_{{\rm P}}t,m).
\label{eq:hat_L_3}
\end{equation}
Equations (\ref{eq:Omega_P_2})-(\ref{eq:hat_L_3}) are the solutions for triaxial free precession adopted in our work. In Appendix \ref{sec:Appendix_B}, we prove that equations (\ref{eq:Omega_P_2})-(\ref{eq:hat_L_3}) are equivalent to equations (\ref{eq:Omega_1})-(\ref{eq:m}) from \cite{landau1960mechanics}.

In the body frame, the dipole moment $\boldsymbol{\mu}$ remains fixed, as illustrated in Figure \ref{Fig_precession}. The unit dipole moment vector components are given by
\begin{equation}
\hat{\mu}_1=\sin\chi\cos\eta,
\label{eq:hat_mu_1}
\end{equation}
\begin{equation}
\hat{\mu}_2=\sin\chi\sin\eta,
\label{eq:hat_mu_2}
\end{equation}
\begin{equation}
\hat{\mu}_3=\cos\chi,
\label{eq:hat_mu_3}
\end{equation}
where $\eta$ is the azimuthal angle, $\chi$ is the polar angle. 

One can find that for a nearly spherical star with very minor deformations, the angular velocity and the angular momentum are almost aligned, with an angle $\hat{\theta} \sim \epsilon_{3}\theta \ll 1$ between them \citep{Jones2001,Levin2020,GaoYong2023}. In the zeroth-order approximation of $\epsilon_3$, the angular velocity is parallel to the angular momentum. In Appendix \ref{sec:Appendix_C}, we provide a detailed explanation of our reasonable approximation that the angular velocity vector is aligned with the angular momentum vector in our work. 
Therefore, the magnetic inclination angle $\alpha$ can be expressed as the angle between the angular momentum $\boldsymbol{L}$ and the dipole moment $\boldsymbol{\mu}$, and it can be defined as
\begin{equation}
\cos\alpha=\hat\mu_1\hat L_1+\hat\mu_2\hat L_2+\hat\mu_3\hat L_3.
\label{eq:cos_alpha}
\end{equation}

By combining equations (\ref{eq:lambda_alpha})-(\ref{eq:L_iso_X}), (\ref{eq:2E})-(\ref{eq:L^2}), and (\ref{eq:epsilon_2})-(\ref{eq:cos_alpha}), the magnetic dipole radiation emitted by a newly born magnetar undergoing triaxial free precession can be derived.

\section{Sample study and light-curve fitting }
\label{sec:Samp_light}
\subsection{GRB samples selection and classification}
\label{subsec:GRB_samp}

\begin{deluxetable*}{lcccccccc}
\label{tab:obser_fit_parameters}
\tablecaption{The table contains the observed characteristics of our Gold, Silver, and Bronze samples, the fitting results of the smooth broken power-law model to the XRT data, and the periodicity found by the LSP algorithm.}
\tablehead{
\multicolumn{1}{l}{GRB} &
\colhead{$T_{90}(\text{s})^{a}$} &
\colhead{$z^{a}$} &
\colhead{$\Gamma^{a}$} &
\colhead{$\theta_{\text{j}}(^\circ)^{b}$} &
\colhead{$\alpha_{1}^{c}$} &
\colhead{$\alpha_{2}^{c}$} &
\colhead{$t_{\text{b}}(\text{s})^{c}$} &
\colhead{$P_{\text{QPO}}(\text{s})^{d}$}
}
\startdata
\object{Gold}  &      \\
\hline
\object{GRB 060202} &  204 & $0.783^{(1)}$ & 2.04 & 5 & 0.288 & 6.546 & 775 & 157  \\
\object{GRB 180620A} & 23.16 & $1.2^{(2)}$ & 1.34 & $3^{(5)}$ & 0.004 & 3.197 & 7853 & $650\pm50^{(6)}$ \\ 
\hline
\object{Silver}  &      \\
\hline
\object{GRB 050730}  &  157 & $3.967^{(3)}$ & 1.58 & 5 & 0.340 & 2.816 & 9217 & 246    \\
\hline
\object{Bronze}  &      \\
\hline
\object{GRB 210610A}  & 8.192 & $3.54^{(4)}$ & 1.86 & 5 & 0.000 & 1.119 & 1154 & None  \\
\enddata
\tablecomments{\\${\rm ^{a}}$ The duration of the prompt emission, redshift, and the X-ray spectral index. \\
${\rm ^{b}}$ The jet opening angle, taken from the literature, or according to \cite{Lv2014}'s assumption, is set to $5\,^\circ$.\\
${\rm ^{c}}$ The decay slopes of the plateau and post-plateau, and the observed end time of the plateau, are obtained from fitting with a smooth broken power law.\\
${\rm ^{d}}$ The periodicity obtained by the LSP algorithm (except for GRB 180620A).\\
Reference. (1)\cite{Butler2007}; (2)\cite{Breeveld2018GCN}; (3)\cite{Chen2005GCN}; (4)\cite{Zhu2021GCN}; (5)\cite{Becerra2019}; (6)\cite{ZouLe2022}.}
\end{deluxetable*}

\begin{figure*}
\centering
\includegraphics[width=0.46\textwidth, angle=0]{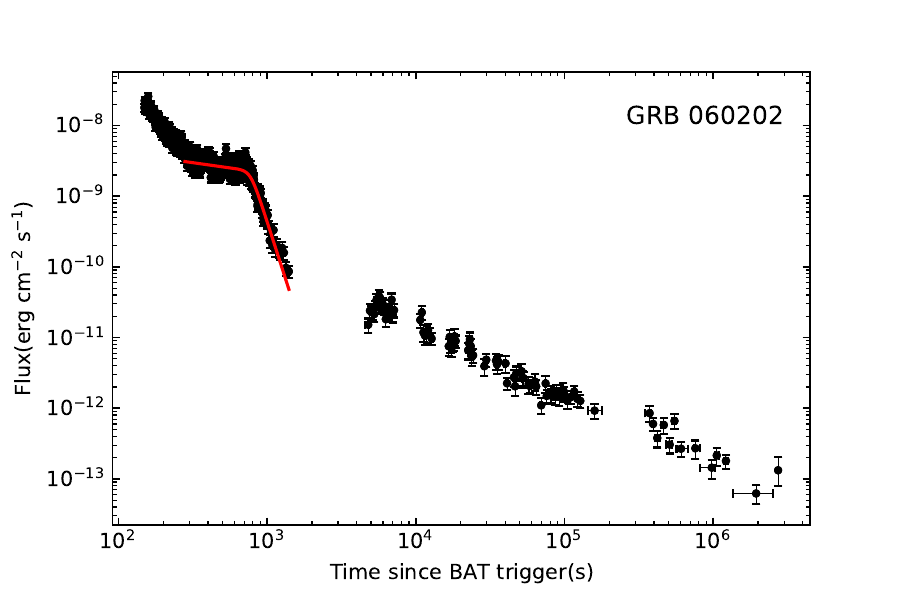}
\includegraphics[width=0.46\textwidth, angle=0]{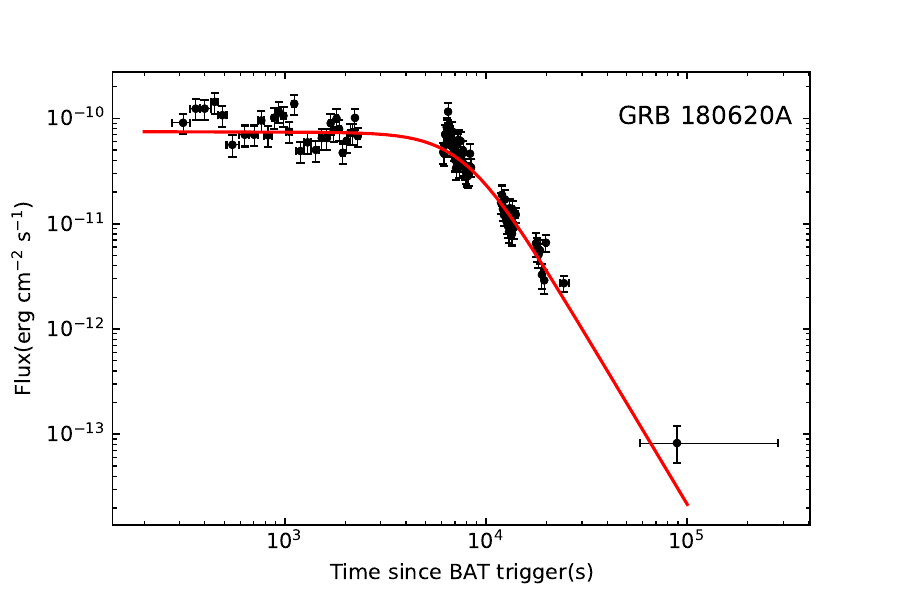}
\includegraphics[width=0.46\textwidth, angle=0]{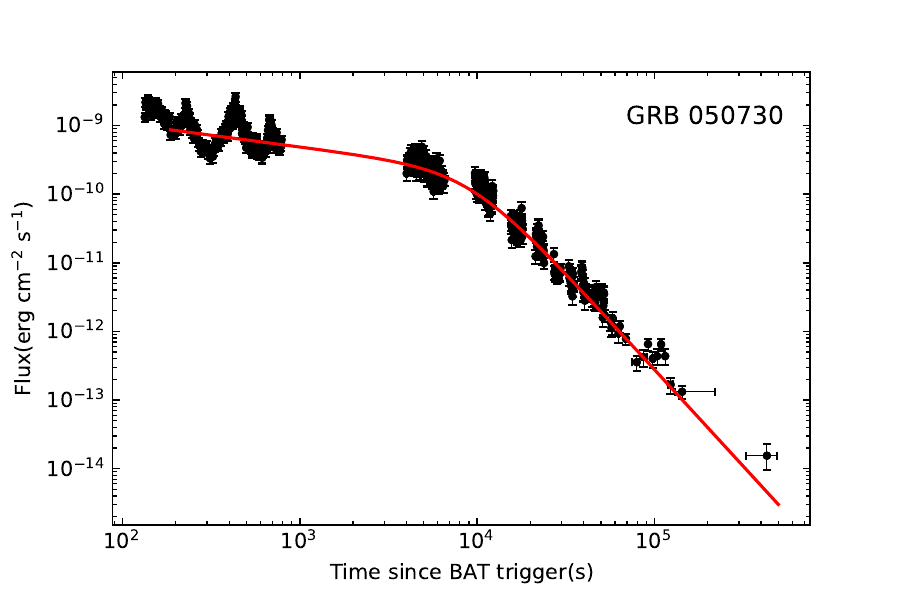}
\includegraphics[width=0.46\textwidth, angle=0]{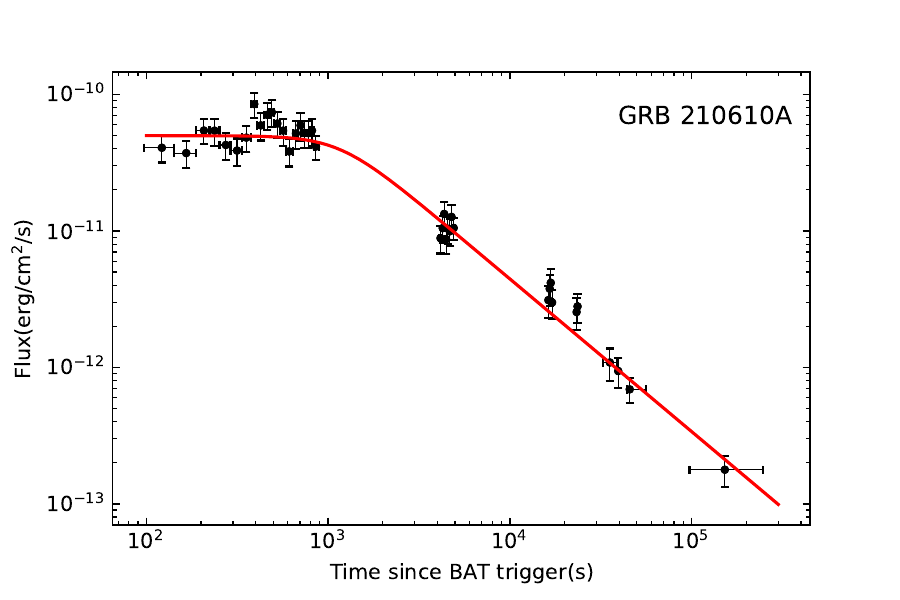}
\caption{The X-ray light curves of GRBs in our sample. GRB 060202 and GRB 180620A are the Gold samples, GRB 050730 is the Silver sample, and GRB 210610A is the Bronze sample. The black data points represent the XRT data of the afterglow, and the red solid curves are the smooth broken power-law model fitted to the XRT data.}
\label{Fig_SMBP}
\end{figure*}

Swift has observed numerous GRBs with X-ray afterglows featuring plateaus \citep{Rowlinson2013,Lv2014,Lv2015,LiLiang2018,Tang&Hang2019,Den&Huang2023}. Among these, some GRBs exhibit characteristics of regular flux variations on the plateau \citep{Suvoro&Kokkotas2021}. Some of these flux variations even display QPO signals \citep{ZouLe2022}. In this paper, our primary focus is on the study of GRBs that exhibit these features. Through visual inspection, we systematically searched the X-ray afterglow data of Swift GRBs observed between May 2005 and November 2023. In this process, we identified four GRBs that exhibited regular flux variations on X-ray afterglow plateaus. These four GRBs were selected as our study sample, their observational properties are given in Table \ref{tab:obser_fit_parameters}. The XRT data we adopted are taken from the website of Swift Burst Analyzer \citep{Evans2009,Evans2010}. To extract features during and after the plateau, we fit the X-ray afterglow light curves with a smooth broken power law
\begin{equation}
F = F_{0}\left[\left(\dfrac{t}{t_{\text{b}}}\right)^{\omega\alpha_{1}}+\left(\dfrac{t}{t_{\text{b}}}\right)^{\omega\alpha_{2}}\right]^{-1/\omega},
\label{eq:F_smooth}
\end{equation}
where $t_{\text{b}}$ represents the observed end time of the plateau, $F_{0}\, \cdot \, 2^{-1/\omega}$ is the flux at this time, $\alpha_{1}$ and $\alpha_{2}$ correspond to the decay index of the flux during the plateau and post-plateau phase, and the sharpness parameter $\omega$ describes the sharpness of the transition between these phases. The fitting results of the light curves for our GRB samples are illustrated in Figure \ref{Fig_SMBP}, and the relevant parameters are provided in Table \ref{tab:obser_fit_parameters}.

In order to investigate whether there is potential periodicity in the flux variations on the plateaus of our GRB samples, we conducted  power-density spectrum (PDS) analysis using the Lomb–Scargle periodogram (LSP) algorithm \citep{Lomb1976,Scargle1982} for each GRB (except for GRB 180620A) during the time intervals where flux variations were present on the plateaus. The obtained results are given in Table \ref{tab:obser_fit_parameters}. Our analysis reveals the presence of QPO signals in GRB 060202 and GRB 050730, with periods of 157\,s and 246\,s, respectively, and the corresponding power values for both signals exceed the false alarm probability (FAP) level of $0.01\%$. For GRB 180620A, we utilized the results obtained by \cite{ZouLe2022}, indicating the presence of a QPO signal with a period of $650\pm50$ s within the interval ${\rm (200, 2300)\,s}$ at a confidence level of $3\sigma$. For the remaining GRBs, no evident periodic signals were identified in their PDS.

For the processing GRB samples, we classified them into three categories, namely \textquotedblleft Gold\textquotedblright, \, \textquotedblleft Silver\textquotedblright, \, and \textquotedblleft Bronze\textquotedblright, \, based on the likelihood of the central engine being a magnetar and the degree of support for the precession (presence of QPO signals). This categorization indicates a gradual weakening in the evidence supporting triaxially precessing  magnetars as the central engines for these samples. 
\begin{itemize}
    \item[1.] Gold Sample: this category includes bursts with a decay slope steeper than 3 in the post-plateau phase, indicating an internal plateau, and QPO signals in the flux variations on the plateau. The samples in this category are GRB 060202 and GRB 180620A.
    \item[2.] Silver Sample: this category comprises bursts with a normal decay phase following the plateau, indicating an external plateau, and QPO signals in the flux variations on the plateau. The only sample in this category is GRB 050730.
    \item[3.] Bronze Sample: this category includes bursts with a normal decay phase following the plateau, and seemingly regular flux variations on the plateau. However, no QPO signals were found in their PDS. GRB 210610A belongs to this category.
\end{itemize}

\subsection{Light-curve fitting}
\label{subsec:Lightcurve}
In Section \ref{sec:radi-prec}, we introduced a model for the magnetic dipole radiation emitted by a triaxially precessing magnetar, which we employed to fit the X-ray afterglow light curve. In our work, we adopt the magnetar radius of $R=10^{6}\,{\rm cm}$. For the Silver and Bronze samples, which exhibit external plateaus, and the Gold samples, which exhibit internal plateaus, the magnetar mass is set as a free parameter. We express $\epsilon_3$ as $\epsilon_3 = \epsilon_2 + \xi$, where $\xi > 0$ since $\epsilon_2$ is less than $\epsilon_3$. The free parameters in our model include the mass of the magnetar M, the radiation efficiency $\eta_{\rm X}$, the surface polar cap magnetic field $B_{\rm p}$, the initial spin period $P_{0}$, the ellipticity $\epsilon_{2}$, the parameter $\xi$ related to $\epsilon_{3}$, the initial wobble angle $\theta_0$, the azimuthal angle of the dipole moment $\eta$, the polar angle of the dipole moment $\chi$, and the parameter $k$. We employed the EMCEE PYTHON package \citep{Foreman2013} to implement the Markov chain Monte Carlo (MCMC) algorithm for fitting the afterglow data and obtaining constraints on the free parameters of the model. To interpret the afterglow data, we sampled the theoretical light curve at the times corresponding to the afterglow data points. We also conducted PDS analysis on the sampled theoretical points of both the Gold and Silver samples using the LSP algorithm. PDSs are computed with uncertainties of the theoretical sampled points taken to be the same as those of the observed points. The periods of the observed and theoretical points obtained in our work (i.e., the peaks of the PDS) all have LSP powers above a FAP level of $0.01\%$. The best-fitting parameters for our Gold, Silver, and Bronze samples are presented in Table \ref{tab:emcee_parameters}, with the corresponding posterior distributions shown in Figure \ref{corner_180620A}-\ref{corner_210610A}. 

\setlength{\tabcolsep}{1pt}
\begin{deluxetable*}{lcccccccccc}
\label{tab:emcee_parameters}
\tablecaption{The fitting parameters of our model to X-ray afterglow data for three categories of GRB samples.}
\tablehead{
\multicolumn{1}{l}{GRB} &
\colhead{$M(\times10^{-2}M_{\odot})$} &
\colhead{$\eta_{\text{X}}(\times10^{-3})$} &
\colhead{$B_{\text{p}}(\times10^{15}\text{G})$} &
\colhead{$P_{0}(\text{ms})$} &
\colhead{$\epsilon_{2}(\times10^{-6})$} &
\colhead{$\xi(\times10^{-6})$} &
\colhead{$\theta_{0}(\text{rad})$} &
\colhead{$\eta(\text{rad)}$} &
\colhead{$\chi(\text{rad)}$} &
\colhead{$k$}
}
\startdata
\object{Gold}  &      \\
\hline
\object{GRB 060202}  & $240.830^{+3.068}_{-1.852}$ & $2.795^{+1.788}_{-0.930}$ & $2.027^{+0.524}_{-0.379}$ & $1.353^{+0.356}_{-0.252}$ & $8.712^{+9.114}_{-5.422}$ & $7.472^{+2.697}_{-2.163}$ & $0.252^{+0.146}_{-0.086}$ & $1.956^{+0.375}_{-0.512}$ & $0.514^{+0.234}_{-0.171}$ & $0.652^{+0.207}_{-0.189}$ \\
\object{GRB 180620A} & $238.451^{+3.169}_{-1.077}$ & $0.407^{+0.666}_{-0.234}$ & $1.106^{+0.664}_{-0.360}$ & $1.844^{+1.134}_{-0.620}$ & $1.538^{+1.724}_{-0.908}$ & $2.894^{+1.738}_{-1.020}$ & $0.458^{+0.145}_{-0.105}$ & $0.440^{+0.187}_{-0.163}$ & $0.550^{+0.180}_{-0.135}$ & $0.821^{+0.110}_{-0.145}$ \\ 
\hline
\object{Silver}  &      \\
\hline
\object{GRB 050730} & $154.751^{+27.553}_{-21.574}$ & $170.638^{+77.673}_{-53.176}$ & $8.489^{+0.972}_{-1.347}$ & $3.976^{+0.519}_{-0.557}$ & $18.124^{+7.981}_{-6.190}$ & $16.510^{+3.934}_{-3.227}$ & $0.500^{+0.115}_{-0.100}$ & $0.004^{+0.006}_{-0.003}$ & $0.811^{+0.134}_{-0.139}$ & $-0.635^{+0.036}_{-0.048}$ \\
\hline
\object{Bronze}  &      \\
\hline
\object{GRB 210610A}  & $146.544^{+27.809}_{-17.083}$ & $49.311^{+20.124}_{-15.616}$ & $8.747^{+0.828}_{-1.229}$ & $6.007^{+0.733}_{-0.808}$ & $21.539^{+18.238}_{-12.251}$ & $25.818^{+8.115}_{-7.111}$ & $0.536^{+0.194}_{-0.187}$ & $0.162^{+0.197}_{-0.109}$ & $0.741^{+0.362}_{-0.350}$ & $-0.423^{+0.082}_{-0.157}$ \\
\enddata
\end{deluxetable*}

\subsubsection{GRB 180620A}
\label{subsubsec:180620A}
On 2018 June 20 at 08:34:58 UTC,  GRB 180620A triggered a response from the Swift/BAT \citep{Evans2018GCN}. GRB 180620A is one of the relatively bright GRBs observed in recent years. It is identified as a long GRB, with a duration of the prompt emission $T_{90} = 23.16 \pm 4.82 \, \text{s}$ in the 15-150 keV band \citep{Stamatikos2018GCN}. It has a redshift limit of 1.2 given by the detection of the Swift/UVOT \citep{Breeveld2018GCN}. The X-ray spectral index on its shallow decay phase is 1.34 \citep{ZouLe2022}. As shown in Figure \ref{Fig_SMBP}, the X-ray afterglow light curve of GRB 180620A exhibits a plateau phase ($\alpha_{1} = 0.004$), followed by a sharp decay phase ($\alpha_{2} = 3.197$), and the observed break time is $t_{\text{b}} = 7853 \, \text{s}$. Multiple significant flux variations are observed on the plateau, in this phase, \cite{ZouLe2022} identified a QPO signal with a period of $\rm{650 \pm 50 \, s}$ in the time interval ${\rm (200, 2300)\,s}$. GRB 180620A is one of the two Gold samples.

\begin{figure*}
\centering
\includegraphics[width=0.46\textwidth, angle=0]{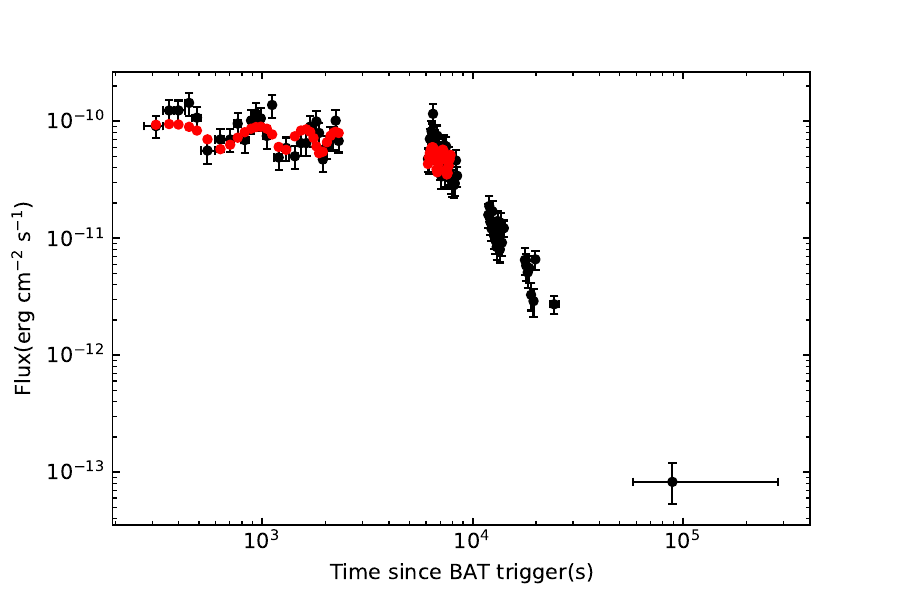}
\includegraphics[width=0.46\textwidth, angle=0]{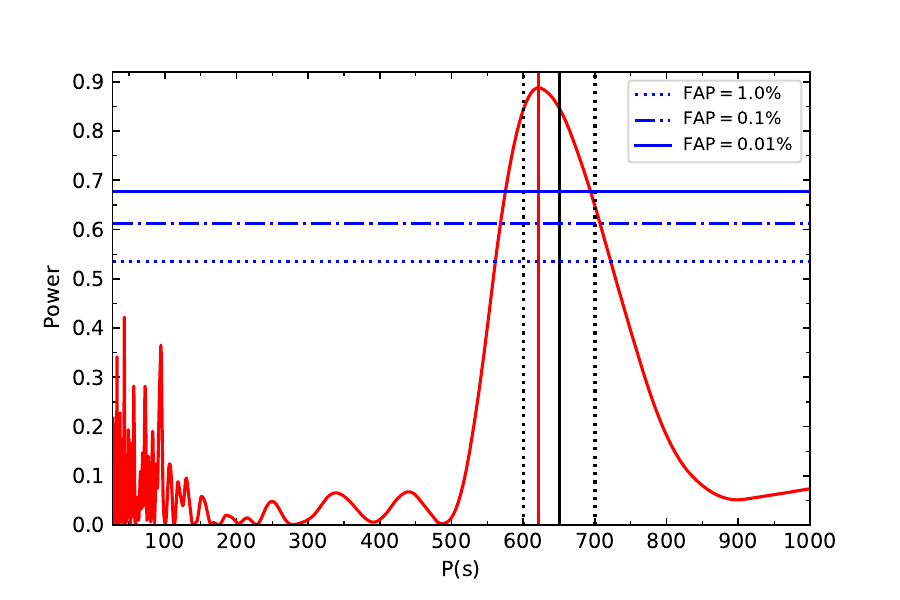}
\caption{Left panel: the sampled theoretical model points (red dots) prior to the break time obtained from the best-fitting model light curve in comparison with the XRT afterglow data (black dots) of GRB 180620A. Right panel: the PDS (red curve) of the theoretical sampled points before ${\rm 2305\,s}$, obtained from LSP. The vertical red solid lines represent the peak of the PDS, corresponding to $P={\rm621\,s}$. The vertical black solid lines and dotted lines represent the period of ${\rm 650\pm50\,s}$  in the time interval ${\rm (200,2300)\, s}$ taken from \cite{ZouLe2022}. The horizontal blue dotted, dashdot, and solid lines represent FAP levels of $1\%$, $0.1\%$, and $0.01\%$, respectively.}
\label{Fig_180620A}
\end{figure*}

We employed our precessing magnetar engine model to fit the XRT data of GRB 180620A prior to the break time (from 312 s to 7853 s, corresponding to the first observed point and the time of the break, respectively). Simultaneously, PDS analysis was conducted on sampled theoretical points in the time interval ${\rm (312,2305)\,s}$. The corresponding theoretical light curve and PDS are illustrated in Figure \ref{Fig_180620A}. Our model provides a good match between the theoretical light curve before the break time and the observed plateau, as well as the flux variations on the plateau. The best-fit values for the mass of the magnetar M, the surface magnetic field $B_{\rm p}$, the initial spin period $P_{0}$, the ellipticity $\epsilon_{2}$, and the parameter $\xi$ related to $\epsilon_{3}$ obtained from the afterglow fitting of GRB 180620A are $M=(238.451^{+3.169}_{-1.077})\times10^{-2}\,M_{\odot}$, $B_{\rm p}=(1.106^{+0.664}_{-0.360})\times10^{15}\,{\rm G}$, $P_0=1.844^{+1.134}_{-0.620}\,{\rm ms}$, $\epsilon_2=(1.538^{+1.724}_{-0.908})\times10^{-6}$, and $\xi=(2.894^{+1.738}_{-1.020})\times10^{-6}$, respectively. We adopt a commonly used explanation for the internal plateau, suggesting that the magnetar may collapse into a black hole after the break time. Therefore, we did not fit the observations after the break time using our model. The PDS of the model points in the time interval ${\rm (312, 2305)\,s}$ peaks at ${\rm 621\,s}$. Our model's period of ${\rm621\,s}$ is in good agreement with the observed period of $\rm{650 \pm 50 \, s}$ reported by \cite{ZouLe2022}. 

\cite{ZouLe2022} also employed a precessing magnetar to explain the X-ray afterglow and periodicity observed in GRB 180620A. Our model differs from \cite{ZouLe2022} in two main aspects. First, \cite{ZouLe2022} utilized a biaxially precessing magnetar as the central engine for GRB, while we opted for a more physically realistic triaxially precessing magnetar. Second, \cite{ZouLe2022} attributed the sharp drop in flux after the break time to a magnetospheric processes, whereas we consider the widely accepted scenario of magnetar collapse to a BH as a more natural explanation.

\begin{figure*}
\centering
\includegraphics[width=0.46\textwidth, angle=0]{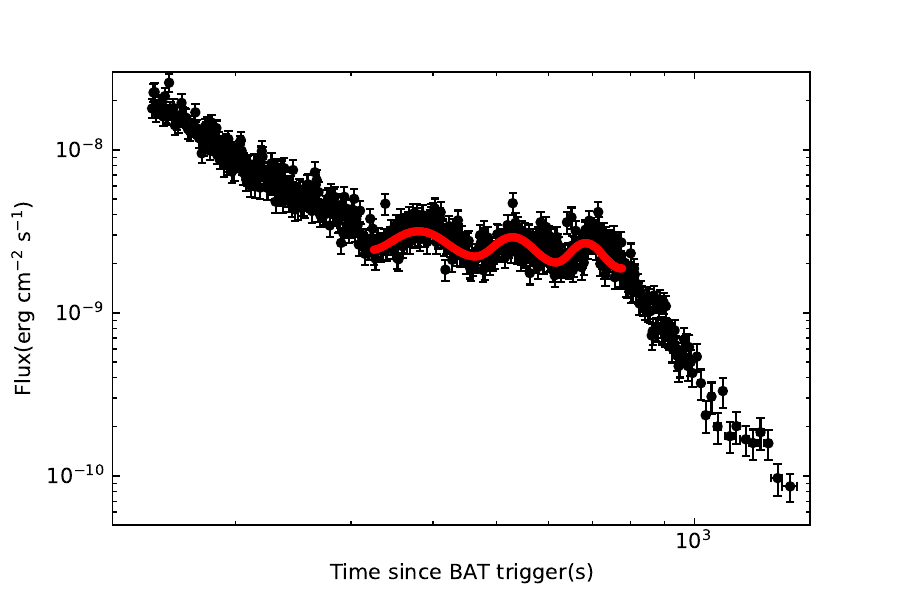}
\includegraphics[width=0.46\textwidth, angle=0]{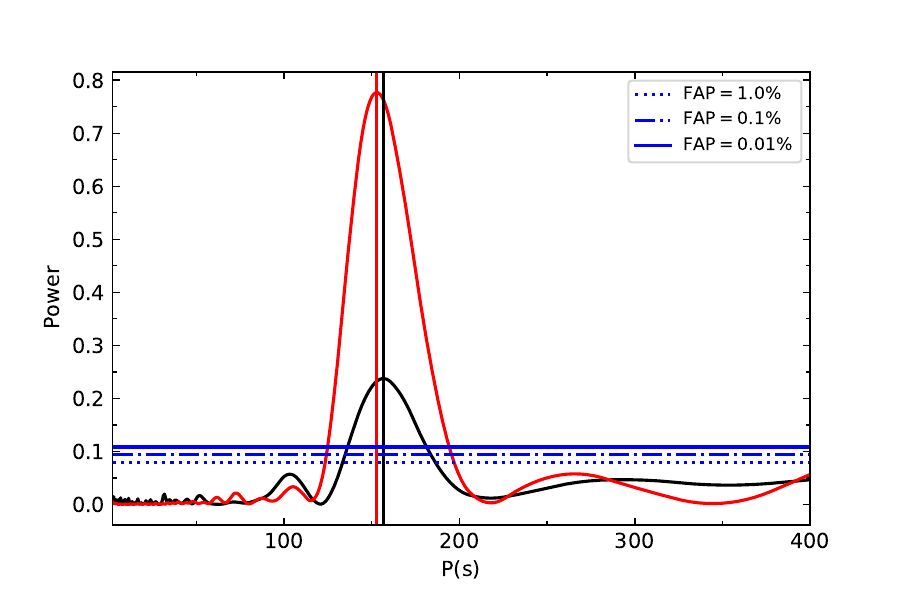}
\caption{Left panel: the sampled points (red dots) from the best-fit model curve during the plateau phase of GRB 060202, and the XRT data (black dots) in time interval ${\rm (149, 1399)\,s}$. Right panel: the PDSs of the observed points (black curve) and the sampled model points (red curve) during the plateau phase obtained by LSP,  with vertical black and red lines representing the corresponding peaks at $P=157\,{\rm s}$ and $P=153\,{\rm s}$, respectively. The same labels as those in Figure \ref{Fig_180620A} are used.}
\label{Fig_060202}
\end{figure*}

\begin{figure*}
\centering
\includegraphics[width=0.46\textwidth, angle=0]{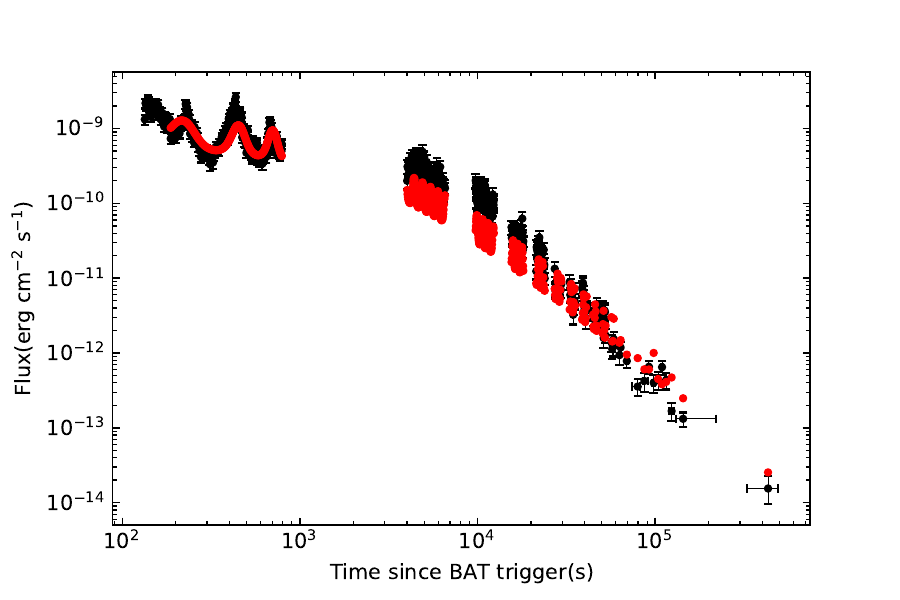}
\includegraphics[width=0.46\textwidth, angle=0]{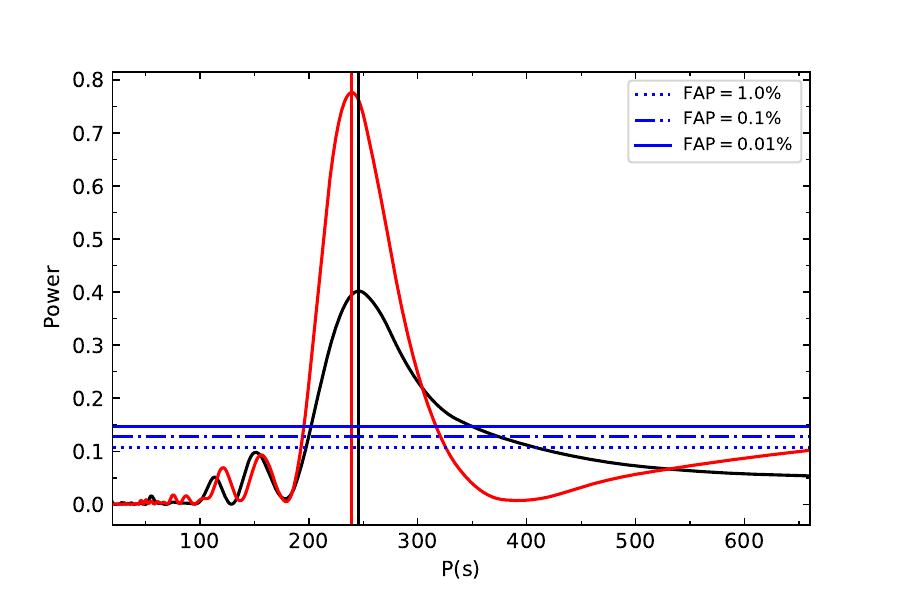}
\caption{Left panel: the sampled X-ray lightcurve (red dots) derived from our best model fit and the XRT data (black dots) of GRB 050730. Right panel: same as the right panel of Figure \ref{Fig_060202}, but for the observations and theoretical points of GRB 050730 in the time interval ${\rm (187, 790)\,s}$. Here, the vertical black and red lines represent $P=246\,{\rm s}$ and $P=239\,{\rm s}$, respectively.}
\label{Fig_050730}
\end{figure*}

\begin{figure*}
\centering
\includegraphics[width=0.46\textwidth, angle=0]{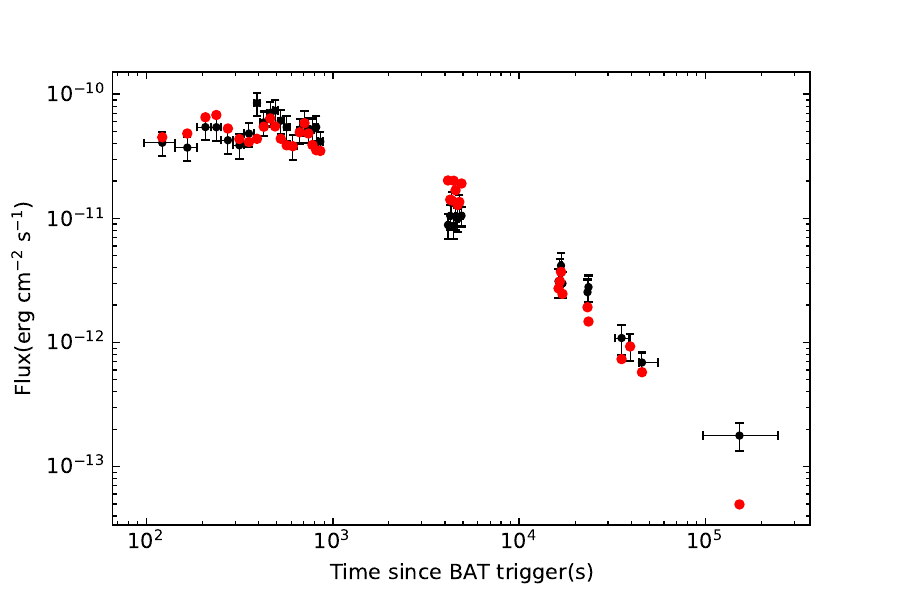}
\includegraphics[width=0.46\textwidth, angle=0]{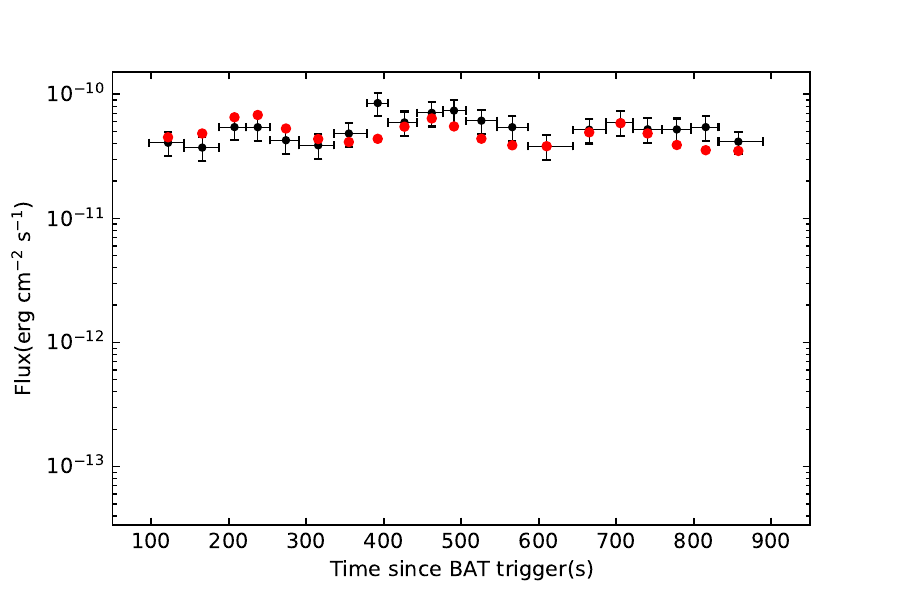}
\caption{Left panel: sampled points (red dots) from the fitted curve and observed afterglow (black dots) of GRB 210610A. Right panel: zoomed-in view of the plateau phase (data before 858 seconds), with the note that the horizontal axis is linear. }
\label{Fig_210610A}
\end{figure*}

\subsubsection{GRB 060202}
\label{subsubsec:060202}
GRB 060202, located at a redshift of 0.783 \citep{Butler2007}, is a long GRB ($T_{90}=204\,\text{s}$) with a long-duration X-ray afterglow. \cite{Lyons2010} noted anomalous regular flux variations in its afterglow during the plateau phase.
GRB 060202 is another Gold sample. As shown in Figure \ref{Fig_SMBP}, the X-ray afterglow of GRB 060202 exhibits the following features: characterized by a plateau phase ($\alpha_1=0.288$), preceded by an initial steep decay phase, followed by a sharp drop phase ($\alpha_2=6.546$), and subsequently a slow decay phase extending up to ${\rm 2.73 \times 10^6\,s}$.

We propose that the initial steep decay phase may arise from curvature effects caused by the cessation of activity of the central engine when the prompt radiation ends. Additionally, similar to GRB 180620A, we attribute the sharp drop after the break time in GRB 060202 to the cessation of energy supply after the collapse of the magnetar into a BH. Thus, we consider only the X-ray afterglow during the plateau phase to be generated by the magnetic dipole radiation of the precessing magnetar. Therefore, we performed MCMC fitting of the afterglow data during the plateau phase (from 325 s to 775 s, corresponding to the beginning and the end of the plateau, respectively) using our model. We also conducted PDS analysis using the LSP algorithm on both the observed points and the sampled theoretical points within the same time interval. The resulting light curve from our model fit and the PDS obtained from LSP are presented in Figure \ref{Fig_060202}. As shown in Figure \ref{Fig_060202}, three regular flux variations are observed in the X-ray afterglow plateau of GRB 060202. Our model successfully reproduces these three flux variations. Moreover, the period obtained from the model's sampled points (153\,s) is consistent with the observed data period (${\rm 157\,s}$). The best-fit values for the mass of the magnetar M, the surface magnetic field $B_{\rm p}$, the initial spin period $P_{0}$, the ellipticity $\epsilon_{2}$, and the parameter $\xi$ related to $\epsilon_{3}$ obtained from the afterglow fitting of GRB 060202 are $M=(240.830^{+3.068}_{-1.852})\times10^{-2}\,M_{\odot}$, $B_{\rm p}=(2.027^{+0.524}_{-0.379})\times10^{15}\,{\rm G}$, $P_0=1.353^{+0.356}_{-0.252}\,{\rm ms}$, $\epsilon_2=(8.712^{+9.114}_{-5.422})\times10^{-6}$, and $\xi=(7.472^{+2.697}_{-2.163})\times10^{-6}$, respectively.  

\subsubsection{GRB 050730}
\label{subsubsec:050730}
GRB 050730, the only Silver sample, is a weak burst detected by Swift/BAT at 19:58:23 UT on July 30, 2005 \citep{Holland2005GCN}. The long GRB 050730, with a prompt emission duration of $T_{90}=157\,\text{s}$, has a high redshift of $z=3.967$ \citep{Chen2005GCN}. The X-ray afterglow of GRB 050730 is characterized by a plateau phase ($\alpha_1=0.340$) followed by a normal decay phase ($\alpha_2=2.816$), but its most prominent feature is the presence of three consecutive flux variations on the plateau. \cite{ZhengTianCi2021} discovered that the time intervals between the peaks of the three flux variations on the plateau are quite similar.

We employed our model to fit the X-ray afterglow of GRB 050730 after 187 seconds, and the resulting fitting curve is shown in the left panel of Figure \ref{Fig_050730}. The declining afterglow observed before 187 seconds may represent the end of the initial steep decay phase commonly observed in X-ray afterglows. The three consecutive flux variations on the plateau derived from the model are in excellent agreement with the three observed flux variations. The best-fit values for the mass of the magnetar M, the surface magnetic field $B_{\rm p}$, the initial spin period $P_{0}$, the ellipticity $\epsilon_{2}$, and the parameter $\xi$ related to $\epsilon_{3}$ obtained from the afterglow fitting of GRB 050730 are $M=(154.751^{+27.553}_{-21.574})\times10^{-2}\,M_{\odot}$, $B_{\rm p}=(8.489^{+0.972}_{-1.347})\times10^{15}\,{\rm G}$, $P_0=3.976^{+0.519}_{-0.557}\,{\rm ms}$, $\epsilon_2=(18.124^{+7.981}_{-6.190})\times10^{-6}$, and $\xi=(16.510^{+3.934}_{-3.227})\times10^{-6}$, respectively. We performed PDS analysis using the LSP on the three flux variations from both the model and observations (spanning from 187\,s to 790\,s), and the results are shown in the right panel of Figure \ref{Fig_050730}. The period obtained from the model ($239\,{\rm s}$) matches that of the observed data (${\rm 246\,s}$) and is roughly consistent with the time intervals between the peaks of the three flux variations. \cite{ZhengTianCi2021} proposed that the three flux variations of GRB 050730 are generated by the magnetar accreting surrounding material three times, driving the jet formation. However, they did not explain how the magnetar produces three accretions with identical time intervals, leading to the periodicity observed in the three consecutive flux variations.

\subsubsection{GRB 210610A}
\label{subsubsec:210610A}
GRB 210610A is a long GRB with $T_{90}=8.192\,\text{s}$ and a redshift of $z=3.54$ \citep{Zhu2021GCN}, belonging to our Bronze sample. As shown in Figure \ref{Fig_SMBP}, its X-ray afterglow is composed of a plateau phase with flux variations and a subsequent slow decay ($\alpha_2=1.119$). Figure \ref{Fig_210610A} presents the fitted curve of our model for the X-ray afterglow of GRB 210610A. As shown in the left panel of Figure \ref{Fig_210610A}, the light curve from our model reproduces the overall features of the observed afterglow. In the right panel of Figure \ref{Fig_210610A}, both the observed and theoretical afterglows on the plateau exhibit seemingly regular flux variations. However, the PDS obtained using the LSP for the plateau phase did not reveal credible periodicity. The best-fit values for the mass of the magnetar M, the surface magnetic field $B_{\rm p}$, the initial spin period $P_{0}$, the ellipticity $\epsilon_{2}$, and the parameter $\xi$ related to $\epsilon_{3}$ obtained from the afterglow fitting of GRB 210610A are $M=(146.544^{+27.809}_{-17.083})\times10^{-2}\,M_{\odot}$, $B_{\rm p}=(8.747^{+0.828}_{-1.229})\times10^{15}\,{\rm G}$, $P_0=6.007^{+0.733}_{-0.808}\,{\rm ms}$, $\epsilon_2=(21.539^{+18.238}_{-12.251})\times10^{-6}$, and $\xi=(25.818^{+8.115}_{-7.111})\times10^{-6}$, respectively.

\subsection{The collapse time and the precession period}
\label{subsec:collapse_period}
\subsubsection{The collapse time for the Gold samples GRB 060202 and GRB 180620A}
\label{subsubsec:collapse_time}
Long GRBs are produced during the core collapse of massive stars \citep{Woosley1993,MacFadyen&Woosley1999}. Some long GRBs exhibit plateau features in their X-ray afterglows, which can be explained by the presence of a magnetar as the central engine \citep{Lv2014}. If the X-ray afterglow plateau of a long GRB is followed by a decay with a slope shallower than $-3$, it is referred to as an external plateau \citep{Lv2015}. This corresponds to the Silver sample GRB 050730 and the Bronze sample GRB 210610A in our work. Conversely, if the X-ray afterglow plateau is followed by a decay with a slope steeper than $-3$, it is referred to as an internal plateau \citep{Troja2007,Rowlinson2010}. This corresponds to the Gold samples GRB 060202 and GRB 180620A in our work. The external plateau of a long GRB can be explained by the long-lived newborn magnetar injecting its rotational energy into an external shock via magnetic dipole radiation, with the end of the plateau corresponding to the spin-down timescale of the newborn magnetar \citep{Lv2014}. The internal plateau of a long GRB can be explained by internal dissipation within the magnetar wind of a supra-massive magnetar that can only survive for a limited time, with the end time of the plateau corresponding  to the collapse time of the supra-massive magnetar into a black hole \citep{Troja2007}. Specifically, the observed collapse time of the supra-massive magnetar is $t_{\rm col} = t_{\rm b}/(1+z)$.

If the mass of a newborn magnetar formed after the collapse of a massive star is less than the maximum mass of a non-rotating NS $M_{\rm TOV}$, then the newborn magnetar can exist stably for a long time. This scenario corresponds to the external plateaus observed in the Silver and Bronze samples in our study. Therefore, in our work, we set the mass of the newborn magnetar for the Silver and Bronze samples as a free parameter less than $M_{\rm TOV}$. If the mass of a newborn magnetar formed after the collapse of a massive star exceeds $M_{\rm TOV}$, then the newborn magnetar is a supra-massive magnetar, which can only survive for a limited time. Due to magnetic dipole spin-down, the supra-massive newborn magnetar will eventually collapse into a black hole. This scenario corresponds to the internal plateaus observed in the Gold sample in our work. The survival time of a supra-massive newborn magnetar, or the collapse time $T_{\rm col}$, depends not only on the magnetic field, initial spin period, and the equation of state (EOS) of the NS, but also on the mass of the newborn magnetar. Therefore, in our work, we set the mass of the newborn magnetar for the Gold sample as a free parameter.

For a specific EOS, the maximum gravitational mass ($M_{\rm max}$) is a function of the spin period, which can be expressed as \citep{Lyford2003,Lasky2014}
\begin{equation}
    M_{\rm max} = M_{\rm TOV}(1+\hat{\alpha}P^{\hat{\beta}}),
\label{eq:M_max}
\end{equation}
where the values of the parameters $\hat{\alpha}$ and $\hat{\beta}$ depend on the specific EOS. As the magnetar spins down due to magnetic dipole radiation, $M_{\rm max}$ gradually decreases. When $M_{\rm max}$ decreases to equal the mass of the magnetar $M$, the centrifugal force can no longer support the magnetar, and it will collapse into a black hole.

When the spin-down of the magnetar is due to magnetic dipole radiation, the evolution of the spinning period is expressed as \citep{Lv2015}
\begin{equation}
    P(t) = P_{0}\left(1+\dfrac{4\pi^{2}}{3c^{3}}\dfrac{B^{2}_{\rm p}R^{6}}{IP^{2}_{0}}t\right)^{1/2}.
\label{eq:Pt_P0}
\end{equation}
Substituting equation (\ref{eq:Pt_P0}) into equation (\ref{eq:M_max}), the theoretical collapse time $T_{\rm col}$ for a supra-massive magnetar collapsing into a black hole can be obtained and expressed as \citep{Lasky2014,Lv2015,LiLong2024}
\begin{equation}
    T_{\rm col} = \dfrac{3c^{3}I}{4\pi^{2}B^{2}_{\rm p}R^{6}}\left[\left(\dfrac{M-M_{\rm TOV}}{\hat{\alpha}M_{\rm TOV}}\right)^{2/\hat{\beta}}-P^{2}_{0}\right].
\label{eq:T_col}
\end{equation}
The works of \cite{Lasky2014} and \cite{Lv2015} support GM1 \citep{Glendenning1991} as the EOS for NSs. In our work, we choose GM1 as the EOS for NSs, with corresponding parameters $M_{\rm TOV} = 2.37\,M_{\odot}$, $\hat{\alpha}=1.58\times10^{-10}\,s^{-\hat{\beta}}$, and $\hat{\beta} = -2.84$ \citep{Lasky2014,Ravi&Lasky2014,Lv2015}. By substituting the values of $M$, $B_{\rm p}$, and $P_0$ (see Table \ref{tab:emcee_parameters}) obtained from fitting the X-ray afterglow light curves into equation (\ref{eq:T_col}), the theoretical collapse time for the supra-massive magnetars in the Gold samples GRB 060202 and GRB 180620A can be derived. For GRB 060202, the theoretical collapse time $T_{\rm col} = 431\,{\rm s}$ is consistent with the observed collapse time $t_{\rm col} = 435\,{\rm s}$. Similarly, for GRB 180620A, the theoretical collapse time $T_{\rm col} = 3549\,{\rm s}$ is also consistent with the observed collapse time $t_{\rm col} = 3570\,{\rm s}$.

\subsubsection{QPOs and the precession period}
\label{subsubsec:precession_period}
We identified QPO signals in the flux variations on the afterglow plateaus of the Gold samples GRB 060202 and GRB 180620A, as well as on the afterglow plateau of the Silver sample GRB 050730. For a newborn magnetar with spin-down controlled by magnetic dipole radiation, the evolution of the spin frequency $\Omega$ follows a broken power law as described by equation (\ref{eq:Omega}). The evolution of the precession frequency $\Omega_{\rm P}$ is related to $\Omega$. Before the spin-down timescale $\tau_{\rm sd}$, $\Omega$ remains nearly constant, and thus $\Omega_{\rm P}$ also remains nearly constant. After $\tau_{\rm sd}$, $\Omega$ decays following a power law, and $\Omega_{\rm P}$ also decays following a power law. 

For the Gold samples GRB 060202 and GRB 180620A, the end of the plateau corresponds to the collapse of the magnetar into a BH, and here the collapse time $T_{\rm col}$ of the magnetar is less than $\tau_{\rm sd}$. During the afterglow plateau phase of the Gold samples GRB 060202 and GRB 180620A, both $\Omega$ and $\Omega_{\rm P}$ remain nearly constant. Therefore, the periods inferred from the afterglow plateaus of GRB 060202 and GRB 180620A remain nearly constant. After the end of the afterglow plateau, the magnetar may collapse into a BH, and the observed afterglow data are no longer produced by our precessing magnetar model.

For the Silver sample GRB 050730, before $\tau_{\rm sd}$, both $\Omega$ and $\Omega_{\rm P}$ do not undergo significant evolution, and thus QPO signals can be observed. After $\tau_{\rm sd}$, $\Omega$ decays following a power law, and thus $\Omega_{\rm P}$ also decays following a power law. The rapid decay of $\Omega_{\rm P}$ further leads to a gradual weakening of the flux variations \citep{Suvorov&Kokkotas2020}. The rapid decay of $\Omega_{\rm P}$ and the continuous weakening of the flux variations make it difficult to observe QPO signals after $\tau_{\rm sd}$. This may explain why we did not detect QPO signals in the afterglow data of GRB 050730 after $\tau_{\rm sd}$. \cite{ZouLe2022} observed QPO signals in the flux variations on the afterglow plateau of the GRB object they studied, but did not observe such signals in the post-plateau data. They suggested that this is because $\Omega_{\rm P}$ did not undergo significant evolution during the plateau phase, allowing the QPO signals to be detected, while in the later stages, the rapid decay of $\Omega_{\rm P}$ led to the absence of detectable QPO signals. This might be the same reason that we observed QPO signals during the afterglow plateau of the Silver sample GRB 050730, but did not detect QPO signals in the post-plateau data. The evolution of $\Omega_{\rm P}$ is also related to the ellipticity. \cite{ZouLe2022} noted that if the ellipticity evolves over time and is controlled by starquakes (while the deformation caused by the magnetic field can be considered constant in the early stages), then the evolution of ellipticity is the same as that of $\Omega$. This would lead to a faster decay of $\Omega_{\rm P}$ after $\tau_{\rm sd}$, making it even more difficult to detect QPO signals in the post-plateau data.

For the Bronze sample GRB 210610A, we did not detect QPO signals in the X-ray afterglow. However, the precessing magnetar model successfully reproduces both the flux variations on the plateau and the overall characteristics of the afterglow light curve.

\section{Conclusion and Discussion}
\label{sec:Con}
In this paper, we have investigated four GRBs with regular flux variations on their X-ray afterglow plateaus. These four GRBs were selected through visual inspection from Swift/XRT afterglow data spanning from May 2005 to November 2023. By fitting the X-ray afterglows with a smooth broken power law, we derived the decay slopes for each GRB after its plateau phase. By applying the LSP algorithm to the PDS analysis of flux variations on each GRB's plateau, we determined their respective periodic or non-periodic nature. Based on whether the decay slope following the plateau is steeper than -3, and on the presence or absence of periodicity in the flux variations on the plateau, these four GRBs were classified into three categories: Gold (GRB 060202 and GRB 180620A), Silver (GRB 050730), and Bronze (GRB 210610A). We proposed a model of magnetic dipole radiation emitted by a triaxially precessing magnetar, in which electromagnetic radiation dominates the magnetar's spin-down process. With this model, we employed the MCMC algorithm to fit the X-ray afterglows of the three categories of GRB samples. By sampling the fitting curve at the observed data points, we obtained the model-sampled points. Except for the post-plateau phase of the Gold sample GRBs (GRB 060202 and GRB 180620A), which may be attributed to the collapse of the magnetar into a BH, and the initial segment of the afterglow data for GRB 060202 and GRB 050730, which may be associated with steep decay formed by curvature effects, the model-sampled points reproduce the remaining portions of the X-ray afterglows for the three categories of GRB samples. Conducting PDS analysis on the model-sampled points for the Gold and Silver samples during the plateau intervals with observed flux variations using the LSP algorithm, we found a good match between the periods derived from the model-sampled points and those from the observational data. Additionally, the parameters obtained from our model fitting are mutually consistent and align with the expectations for a newborn magnetar. These results suggest that triaxially precessing magnetars can serve as the central engines for these four GRBs, accounting for the observed X-ray afterglows.

Our work invokes a more physically plausible triaxial precessing magnetar as the central engine, replacing the biaxial precessing magnetar model previously used in GRBs studies (studied by \cite{Suvorov&Kokkotas2020} for GRB 080602 and GRB 090510, \cite{ZouLe2021} for GRB 101225A, and \cite{ZouLe2022} for GRB 180620A). Additionally, our work expands the sample of GRBs with QPOs in their early X-ray afterglows, including GRB 060202 and GRB 050730. While our Bronze sample did not exhibit periodicity in its early X-ray afterglow, the flux variations observed on its plateau and the overall characteristics of the afterglow can also be successfully reproduced by the triaxially precessing magnetar model. \cite{Suvoro&Kokkotas2021} identified 16 short GRBs out of 25 with plateau phases in their X-ray light curves that support a precessing magnetar as the central engine. Our work complements their findings by increasing the number of long GRBs that also support the precessing magnetar as the central engine. Our work, along with previous studies, provides evidence for the early precession of newborn millisecond magnetars formed in massive star collapse events (for long GRBs) or binary neutron star merger events (for short GRBs). Furthermore, our work further reinforces the theoretical interpretation that links the plateaus in X-ray afterglows to energy injection from the magnetar engine \citep{Zhang2001}. Particularly, for our Gold sample, the precessing magnetar model can simultaneously explain the presence of internal plateaus and the periodicity of flux variations on these plateaus.

In our work, we employed a triaxially precessing magnetar model, as described in Section \ref{sec:radi-prec}, to account for the regular flux variations observed on the early X-ray plateau of GRBs and their periodicity, if present. There are several important aspects that deserve to be pointed out. First, the ratio of gravitational wave luminosity to electromagnetic radiation luminosity for the magnetar is $L_{\text{GW}}/L_{\text{EM}}\sim 0.3(\epsilon^{2}_{3,-3}B^{-2}_{\text{P,15}}P^{-2}_{0,-3})$ \citep{Suvoro&Kokkotas2021}. By substituting the parameters obtained from fitting the X-ray afterglows of the four GRBs in our sample, it is found that $L_{\text{GW}}\ll L_{\text{EM}}$ for all the GRBs, consistent with the assumption in our model that the spin-down of the magnetar for the four GRBs is dominated by electromagnetic radiation. Second, prior to the spin-down timescale $\tau_{\text{sd}}$, the evolution of the magnetic inclination angle $\alpha$ is primarily governed by precession \citep{Goldreich1970,Zanazzi2015}. Our work does not take into account the other factors that might influence $\alpha$ in the later stages, which can be explored through Equation (7) in \cite{Goldreich1970} or Equation (56) in \cite{Zanazzi2015}. Third, in our work, we invoked a model of a freely precessing magnetar. In more realistic scenarios, a magnetar may be subject to external torques, such as precession under the influence of near-field torques \citep{Zanazzi2015}, precession driven by both near-field and far-field torques \citep{GaoYong2023}, and precession of magnetars with plasma-filled magnetospheres \citep{Arzamasskiy2015}. Applying these more complex and realistic magnetar precession dynamics to explain the X-ray afterglows of GRBs requires further investigation in the future.

\section*{acknowledgments}
We thank the referee for valuable comments, which have helped us significantly improve our manuscript.
We also thank Yu-Chen Huang, Zi-Bin Zhang, Min-Xuan Cai, Tian-Yu Xia, and Hui Hong for useful discussions. 
We acknowledge the use of the public data from the Swift data archive and the UK Swift Science Data Center.
This work was supported by the National SKA Program of China (grant No. 2020SKA0120302) and the National Natural Science Foundation of China (grant No. 12393812). S.Q.Z. is supported by the starting Foundation of Guangxi University of Science and Technology (grant No. 24Z17). L.L. acknowledges support from the National Natural Science Foundation of China (grant No. 12303050) and the Fundamental Research Funds for the Central Universities.

\bibliographystyle{aasjournal}
\bibliography{refs}

\appendix 
\section{The posterior probability distribution of model parameters obtained by MCMC fit}
\label{sec:Appendix_A}
\begin{figure*}[b]
\centering
\includegraphics[width=0.9\textwidth, angle=0]{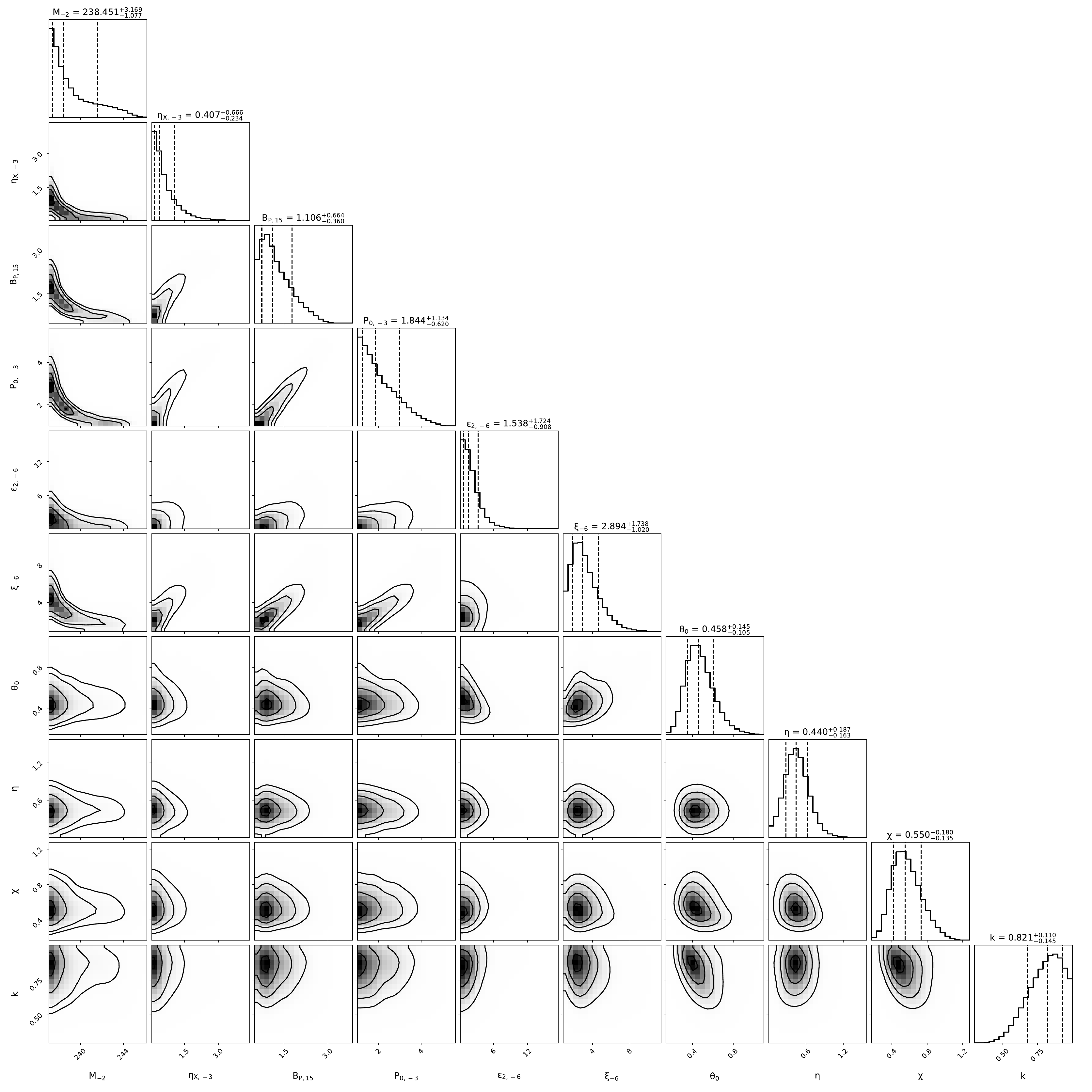}
\caption{The corner plot of the posterior probability distribution of model parameters obtained by MCMC fitting the X-ray afterglow of GRB 180620A with our model. The best-fitting parameters and the corresponding $1\sigma$ uncertainties are depicted in the diagonal histograms with black dashed lines.}
\label{corner_180620A}
\end{figure*}

\begin{figure*}
\centering
\includegraphics[width=\textwidth, angle=0]{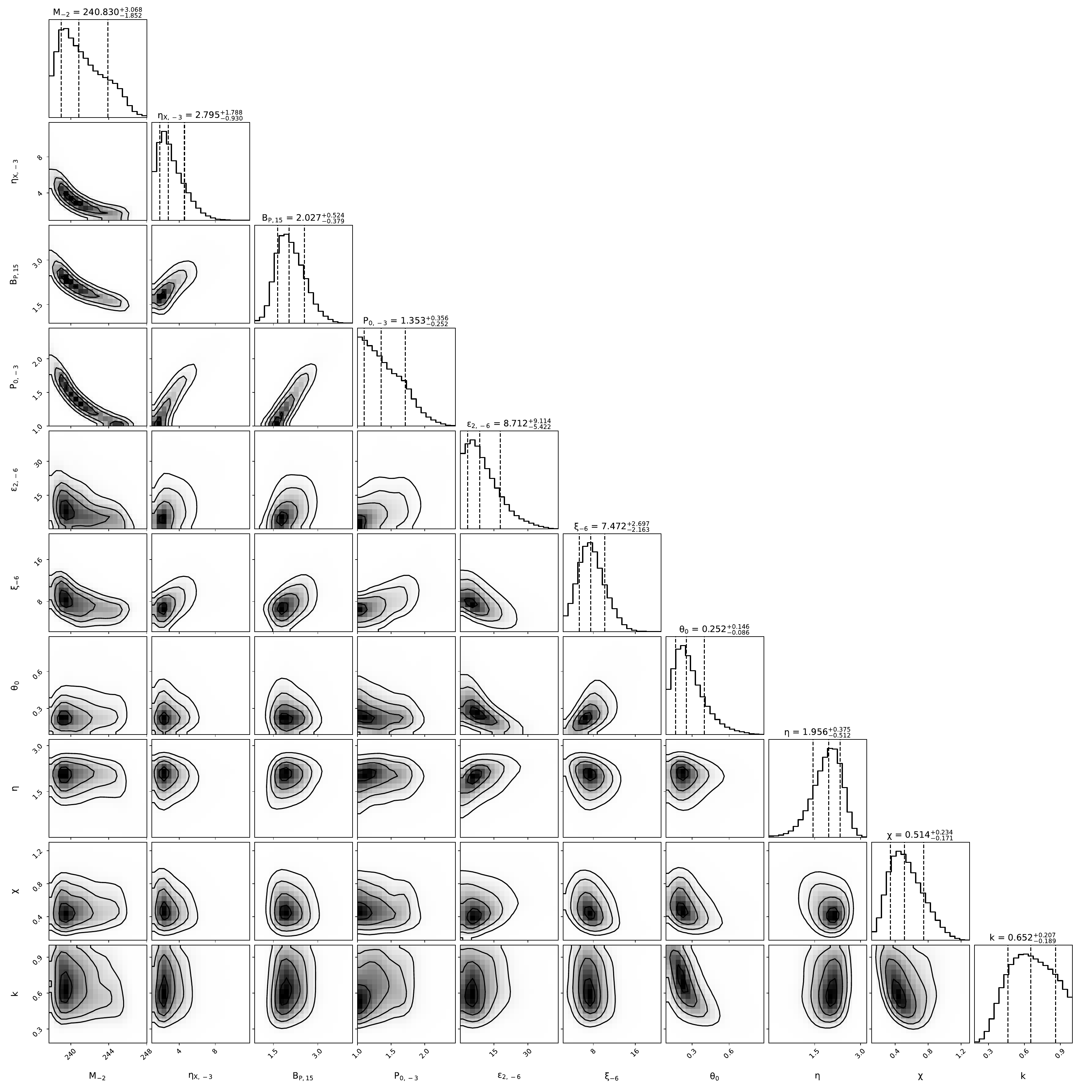}
\caption{Same as Figure \ref{corner_180620A} but for GRB 060202.}
\label{corner_060602}
\end{figure*}

\begin{figure*}
\centering
\includegraphics[width=\textwidth, angle=0]{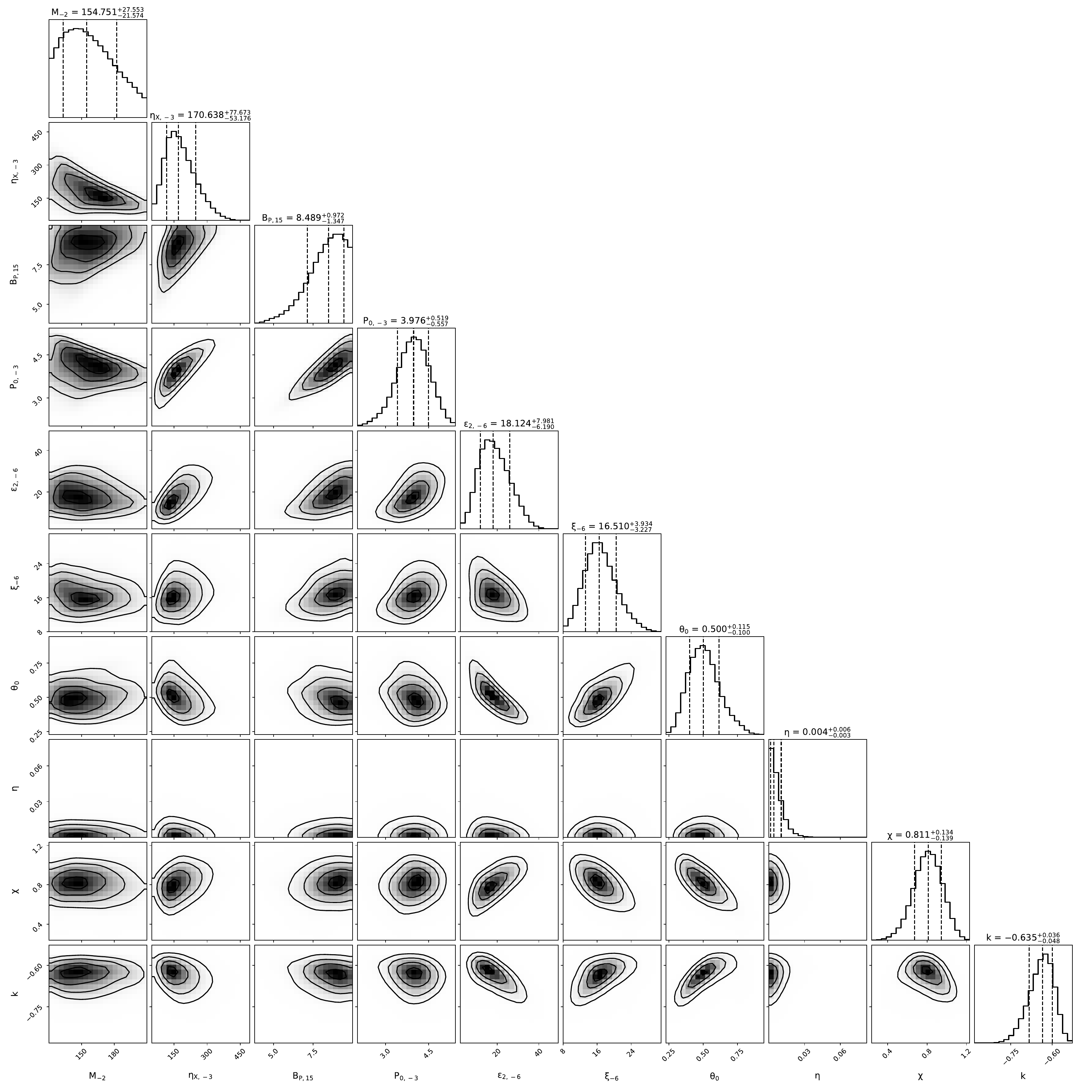}
\caption{Same as Figure \ref{corner_180620A} but for GRB 050730.}
\label{corner_050730}
\end{figure*}

\begin{figure*}
\centering
\includegraphics[width=\textwidth, angle=0]{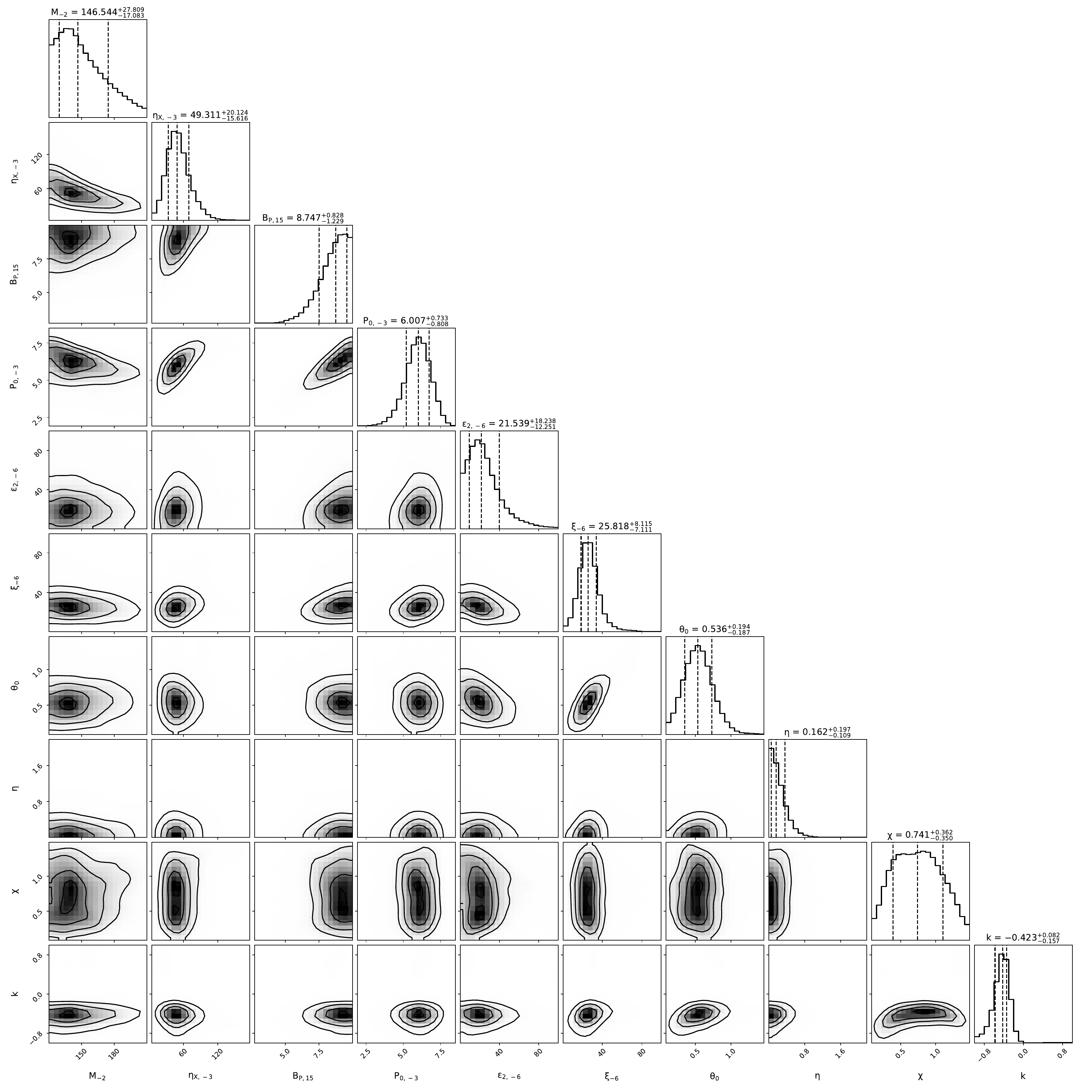}
\caption{Same as Figure \ref{corner_180620A} but for GRB 210610A.}
\label{corner_210610A}
\end{figure*}

In Figures \ref{corner_180620A}-\ref{corner_210610A}, we present the corner plots obtained from the fits of GRB 180620A, GRB 060202, GRB 050730, and GRB 210610A, respectively.

\clearpage
\section{Equivalence of Two Sets of Solutions for Triaxial Free Precession}
\label{sec:Appendix_B}
In this section, we will prove the equivalence of the two sets of solutions for triaxial free precession presented in Section \ref{subsec:precession}. Specifically, we will prove that equations (\ref{eq:Omega_1})-(\ref{eq:m}) and equations (\ref{eq:Omega_P_2})-(\ref{eq:hat_L_3}) are equivalent, with the former derived from \cite{landau1960mechanics} and the latter from \cite{GaoYong2023}.

In this paper, we adopt the same initial conditions as \cite{landau1960mechanics} and \cite{GaoYong2023}, i.e., at the initial time $t=0$, $\Omega_{2}=0$. As in \cite{GaoYong2023}, the wobble angle $\theta$ is taken as $\theta_{0}$ at the initial time $t=0$. Therefore, at $t=0$, we have $L_{1}=I_{1}\Omega_{1}=L\sin\theta_{0}$, $L_{2}=I_{2}\Omega_{2}=0$, and $L_{3}=I_{3}\Omega_{3}=L\cos\theta_{0}$. At $t=0$, from equation (\ref{eq:2E}), we obtain that
\begin{equation}
2E = I_{1}\Omega^{2}_{1}+I_{2}\Omega^{2}_{2}+I_{3}\Omega^{2}_{3} = \dfrac{L^{2}_{1}}{I_{1}}+\dfrac{L^{2}_{2}}{I_{2}}+\dfrac{L^{2}_{3}}{I_{3}} = \dfrac{L^{2}\sin^{2}\theta_{0}}{I_{1}}+\dfrac{L^{2}\cos^{2}\theta_{0}}{I_{3}}.
\label{eq:2E_2}
\end{equation}
For triaxial free precession, both energy and angular momentum are conserved, so $E$ and $L$ remain constant. From equation (\ref{eq:Omega_1}), one can obtain that
\begin{equation}
\hat{L}_{1} = \dfrac{L_{1}}{L} = \dfrac{I_{1}\Omega_{1}}{L} = \dfrac{I_{1}}{L}\sqrt{\frac{2EI_3-L^2}{I_1(I_3-I_1)}}\,{\rm cn}(\tau,m).
\label{eq:hat_L_1_3}
\end{equation}
Substituting equation (\ref{eq:2E_2}) into equation (\ref{eq:hat_L_1_3}), we obtain that 
\begin{eqnarray}
\hat{L}_{1} &=& \dfrac{I_{1}}{L}\sqrt{\frac{(I_3/I_1)L^{2}\sin^{2}\theta_{0}+L^{2}\cos^{2}\theta_{0}-L^2}{I_1(I_3-I_1)}}\,{\rm cn}(\tau,m) \nonumber \\
&=& \sqrt{\frac{I_{3}L^{2}\sin^{2}\theta_{0}+I_{1}L^{2}\cos^{2}\theta_{0}-I_{1}L^2}{L^{2}(I_3-I_1)}}\,{\rm cn}(\tau,m) \nonumber \\
&=& \sqrt{\frac{L^{2}\sin^{2}\theta_{0}(I_{3}-I_{1})}{L^{2}(I_3-I_1)}}\,{\rm cn}(\tau,m) = \sin\theta_{0}\,{\rm cn}(\tau,m).
\label{eq:hat_L_1_4}
\end{eqnarray}
Similarly, combining equation (\ref{eq:Omega_3}) and equation (\ref{eq:2E_2}), we can obtain 
\begin{eqnarray}
\hat{L}_{3} = \dfrac{L_{3}}{L} = \cos\theta = \dfrac{I_{3}\Omega_{3}}{L} &=& \dfrac{I_{3}}{L}\sqrt{\frac{L^2-2EI_1}{I_3(I_3-I_1)}}\,{\rm dn}(\tau,m) \nonumber \\
&=& \dfrac{I_{3}}{L}\sqrt{\frac{L^2-L^{2}\sin^{2}\theta_{0}-(I_{1}/I_{3})L^{2}\cos^{2}\theta_{0}}{I_3(I_3-I_1)}}\,{\rm dn}(\tau,m)
\nonumber \\
&=& \sqrt{\frac{I_{3}L^2-I_{3}L^{2}\sin^{2}\theta_{0}-I_{1}L^{2}\cos^{2}\theta_{0}}{L^{2}(I_3-I_1)}}\,{\rm dn}(\tau,m) \nonumber \\
&=& \sqrt{\frac{L^{2}\cos^{2}\theta_{0}(I_{3}-I_{1})}{L^{2}(I_3-I_1)}}\,{\rm dn}(\tau,m) = \cos\theta_{0}\,{\rm dn}(\tau,m).
\label{eq:hat_L_3_3}
\end{eqnarray}
From equations (\ref{eq:hat_L_1_3}), (\ref{eq:hat_L_1_4}), and (\ref{eq:hat_L_3_3}), one can find that 
\begin{equation}
\sin{\theta_0} = \dfrac{I_{1}}{L}\sqrt{\frac{2EI_3-L^2}{I_1(I_3-I_1)}} = \sqrt{\frac{I_{1}(2EI_3-L^2)}{L^{2}(I_3-I_1)}},
\label{eq:sin_theta_0}
\end{equation}
and
\begin{equation}
\cos{\theta_0} = \dfrac{I_{3}}{L}\sqrt{\frac{L^2-2EI_1}{I_3(I_3-I_1)}} = \sqrt{\frac{I_{3}(L^2-2EI_1)}{L^{2}(I_3-I_1)}}.
\label{eq:cos_theta_0}
\end{equation}
Combining equation (\ref{eq:Omega_2}) and equation (\ref{eq:2E_2}), one can obtain that
\begin{eqnarray}
    \hat{L}_{2} = \dfrac{L_{2}}{L} = \dfrac{I_{2}\Omega_{2}}{L} &=& \dfrac{I_2}{L}\sqrt{\frac{2EI_3-L^2}{I_2(I_3-I_2)}}\,{\rm sn}(\tau,m) \nonumber \\ 
    &=& \sqrt{\frac{I_{2}(2EI_3-L^2)}{L^{2}(I_3-I_2)}}\,{\rm sn}(\tau,m) \nonumber \\ 
    &=& \sqrt{\frac{I_{1}(2EI_3-L^2)}{L^{2}(I_3-I_1)}}\cdot\sqrt{\dfrac{I_{2}(I_{3}-I_{1})}{I_{1}(I_{3}-I_{2})}}\,{\rm sn}(\tau,m) \nonumber \\
    &=& \sqrt{\frac{I_{1}(2EI_3-L^2)}{L^{2}(I_3-I_1)}}\cdot\sqrt{1+\dfrac{I_{3}(I_{2}-I_{1})}{I_{1}(I_{3}-I_{2})}}\,{\rm sn}(\tau,m) \nonumber \\
    &=& \sin{\theta_{0}}\,\sqrt{1+\delta}\,{\rm sn}(\tau,m),
\label{eq:hat_L_2_3}
\end{eqnarray}
where the expressions for $\sin{\theta_{0}}$ and $\delta$ are derived from equation (\ref{eq:sin_theta_0}) and equation (\ref{eq:delta}), respectively. 
To prove that equations (\ref{eq:Omega_1})-(\ref{eq:m}) and equations (\ref{eq:Omega_P_2})-(\ref{eq:hat_L_3}) are equivalent, it is also necessary to prove that equation (\ref{eq:Omega_P_2}) is equivalent to equation (\ref{eq:Omega_P}), and that equation (\ref{eq:m_2}) is equivalent to equation (\ref{eq:m}). Substituting equation (\ref{eq:epsilon_3}), equation (\ref{eq:delta}), and equation (\ref{eq:cos_theta_0}) into equation (\ref{eq:Omega_P_2}), one can obtain that
\begin{eqnarray}
    \Omega_{{\rm P}} &=& \dfrac{\epsilon_{3}L\cos\theta_0}{I_{3}\sqrt{1+\delta}} \nonumber \\
    &=& \dfrac{I_{3}-I_{1}}{I_{1}}\cdot\dfrac{L}{I_{3}}\cdot\sqrt{\frac{I_{3}(L^2-2EI_1)}{L^{2}(I_3-I_1)}}\cdot\left[1+\dfrac{I_{3}(I_{2}-I_{1})}{I_{1}(I_{3}-I_{2})}\right]^{-\dfrac{1}{2}} \nonumber \\
    &=& \dfrac{I_{3}-I_{1}}{I_{1}}\cdot\dfrac{L}{I_{3}}\cdot\sqrt{\frac{I_{3}(L^2-2EI_1)}{L^{2}(I_3-I_1)}}\cdot\sqrt{\dfrac{I_{1}(I_{3}-I_{2})}{I_{2}(I_{3}-I_{1})}} \nonumber \\
    &=& \sqrt{\frac{(I_3-I_2)(L^2-2EI_1)}{I_1I_2I_3}}.
\label{eq:Omega_P_4}
\end{eqnarray}
Thus, equation (\ref{eq:Omega_P_2}) is equivalent to equation (\ref{eq:Omega_P}). Substituting equation (\ref{eq:delta}), equation (\ref{eq:sin_theta_0}), and equation (\ref{eq:cos_theta_0}) into equation (\ref{eq:m_2}), one can obtain that 
\begin{eqnarray}
    m = \delta\tan^{2}\theta_{0} &=& \delta\,\dfrac{\sin^{2}\theta_{0}}{\cos^{2}\theta_{0}} \nonumber \\
    &=& \dfrac{I_{3}(I_{2}-I_{1})}{I_{1}(I_{3}-I_{2})}\cdot\frac{I_{1}(2EI_3-L^2)}{L^{2}(I_3-I_1)}\cdot\left[\frac{I_{3}(L^2-2EI_1)}{L^{2}(I_3-I_1)}\right]^{-1} \nonumber \\
    &=& \dfrac{I_{3}(I_{2}-I_{1})}{I_{1}(I_{3}-I_{2})}\cdot\frac{I_{1}(2EI_3-L^2)}{L^{2}(I_3-I_1)}\cdot\dfrac{L^{2}(I_{3}-I_{1})}{I_{3}(L^{2}-2EI_{1})} \nonumber \\
    &=& \dfrac{(I_{2}-I_{1})(2EI_{3}-L^{2})}{(I_{3}-I_{2})(L^{2}-2EI_{1})}.
\label{eq:m_3}
\end{eqnarray}
Thus, equation (\ref{eq:m_2}) is equivalent to equation (\ref{eq:m}). Combining equations (\ref{eq:2E_2})-(\ref{eq:m_3}), we have proven that equations (\ref{eq:Omega_1})-(\ref{eq:m}) and equations (\ref{eq:Omega_P_2})-(\ref{eq:hat_L_3}) are equivalent, meaning the two sets of solutions for triaxial free precession are equivalent.

\section{Discussion on the relationship between the direction of the angular velocity vector and the direction of the angular momentum vector}
\label{sec:Appendix_C}
In studying the free precession of neutron stars, \cite{Jones2001} found that for a nearly spherical star, the angle $\hat{\theta}\simeq\epsilon_{3}\theta$ between the angular velocity vector and the angular momentum vector is much smaller than the wobble angle $\theta$ (see equation 18 in \citealt{Jones2001}). \cite{GaoYong2023} found that $\hat{\theta}\sim\epsilon_{3}\theta\ll1$, and while studying the triaxial precession of magnetars, they approximated the angular velocity vector as being parallel to the angular momentum vector (see equation 18 and the explanation below equation 18 in \citealt{GaoYong2023}). In their work on explaining the periodicity of FRB 180916.J0158+65 using a triaxial precessing magnetar, \cite{Levin2020} proposed that for a nearly spherical star, since its inertia tensor is nearly equal to a multiple of the unit matrix, the angular velocity vector and the angular momentum vector are almost aligned (see the fifth paragraph of Section 2.1 and Figure 1 in \citealt{Levin2020}).

In this section, we will discuss the relationship between the direction of the angular velocity vector and the direction of the angular momentum vector in the body frame. We find that for objects with $\epsilon_{2}\ll1$ and $\epsilon_{3}\ll1$, it is a reasonable approximation that the angular velocity vector is aligned with the angular momentum vector, which is consistent with the results of \cite{Levin2020} and \cite{GaoYong2023}.

The unit angular momentum vector and the unit angular velocity vector in the body frame are expressed as $\boldsymbol{\hat{L}}=\hat{L}_{1}\boldsymbol{\hat{e}_{1}}+\hat{L}_{2}\boldsymbol{\hat{e}_{2}}+\hat{L}_{3}\boldsymbol{\hat{e}_{3}}$ and $\boldsymbol{\hat{\Omega}}=\hat{\Omega}_{1}\boldsymbol{\hat{e}_{1}}+\hat{\Omega}_{2}\boldsymbol{\hat{e}_{2}}+\hat{\Omega}_{3}\boldsymbol{\hat{e}_{3}}$, respectively. In the body frame, one can obtain that
\begin{equation}
    \hat{\Omega}_{1} = \dfrac{\Omega_{1}}{\Omega}= \dfrac{L_{1}}{I_{1}\Omega}= \dfrac{L\hat{L}_{1}}{I_{1}\Omega} = \dfrac{L}{\Omega}\cdot\dfrac{\hat{L}_{1}}{I_{1}},
\label{eq:hat_Omega_1}
\end{equation}
\begin{equation}
    \hat{\Omega}_{2} = \dfrac{\Omega_{2}}{\Omega}= \dfrac{L_{2}}{I_{2}\Omega}= \dfrac{L\hat{L}_{2}}{I_{2}\Omega} = \dfrac{L}{\Omega}\cdot\dfrac{\hat{L}_{2}}{I_{2}},
\label{eq:hat_Omega_2}
\end{equation}
\begin{equation}
    \hat{\Omega}_{3} = \dfrac{\Omega_{3}}{\Omega}= \dfrac{L_{3}}{I_{3}\Omega}= \dfrac{L\hat{L}_{3}}{I_{3}\Omega} = \dfrac{L}{\Omega}\cdot\dfrac{\hat{L}_{3}}{I_{3}}.
\label{eq:hat_Omega_3}
\end{equation}
According to equation (\ref{eq:hat_Omega_1}), equation (\ref{eq:hat_Omega_2}), and equation (\ref{eq:hat_Omega_3}), the cosine of the angle $\hat{\theta}$ between $\boldsymbol{\hat{L}}$ and $\boldsymbol{\hat{\Omega}}$ can be expressed as
\begin{eqnarray}
    \cos{\hat{\theta}} = \boldsymbol{\hat{L}}\cdot\boldsymbol{\hat{\Omega}} &=& \hat{\Omega}_{1}\hat{L}_{1} + \hat{\Omega}_{2}\hat{L}_{2} + \hat{\Omega}_{3}\hat{L}_{3} \nonumber \\
    &=& \dfrac{L}{\Omega} \cdot \left(\dfrac{\hat{L}^{2}_{1}}{I_{1}} + \dfrac{\hat{L}^{2}_{2}}{I_{2}} + \dfrac{\hat{L}^{2}_{3}}{I_{3}}\right) \nonumber \\
    &=& \dfrac{L}{\Omega I_{1}} \cdot \left(\hat{L}^{2}_{1} + \dfrac{\hat{L}^{2}_{2}}{I_{2}/I_{1}} + \dfrac{\hat{L}^{2}_{3}}{I_{3}/I_{1}}\right) \nonumber \\
    &=& \dfrac{L}{\Omega I_{1}} \cdot \left(\hat{L}^{2}_{1} + \dfrac{\hat{L}^{2}_{2}}{1+\epsilon_{2}} + \dfrac{\hat{L}^{2}_{3}}{1+\epsilon_{3}}\right),
\label{eq:cos_hat_theta}
\end{eqnarray}
where $I_{2}/I_{1}=1+\epsilon_{2}$ and $I_{3}/I_{1}=1+\epsilon_{3}$ come from equation (\ref{eq:I_2}) and equation (\ref{eq:I_3}), respectively. The first term on the right-hand side of equation (\ref{eq:cos_hat_theta}) can be further expressed as
\begin{eqnarray}
    \dfrac{L}{\Omega I_{1}} &=& \dfrac{\sqrt{L^{2}_{1}+L^{2}_{2}+L^{2}_{3}}}{I_{1}\sqrt{\Omega^{2}_{1}+\Omega^{2}_{2}+\Omega^{2}_{3}}} \nonumber \\
    &=& \dfrac{\sqrt{I^{2}_{1}\Omega^{2}_{1}+I^{2}_{2}\Omega^{2}_{2}+I^{2}_{3}\Omega^{2}_{3}}}{I_{1}\sqrt{\Omega^{2}_{1}+\Omega^{2}_{2}+\Omega^{2}_{3}}} \nonumber \\
    &=& \sqrt{\dfrac{\Omega^{2}_{1} + (I_{2}/I_{1})^{2}\Omega^{2}_{2} + (I_{3}/I_{1})^{2}\Omega^{2}_{3}}{\Omega^{2}_{1}+\Omega^{2}_{2}+\Omega^{2}_{3}}} \nonumber \\
    &=& \sqrt{\dfrac{\Omega^{2}_{1} + (1+\epsilon_{2})^{2}\Omega^{2}_{2} + (1+\epsilon_{3})^{2}\Omega^{2}_{3}}{\Omega^{2}_{1}+\Omega^{2}_{2}+\Omega^{2}_{3}}} \nonumber \\
    &=& \sqrt{\dfrac{\Omega^{2}_{1}+\Omega^{2}_{2}+2\epsilon_{2}\Omega^{2}_{2}+\epsilon^{2}_{2}\Omega^{2}_{2}+\Omega^{2}_{3}+2\epsilon_{3}\Omega^{2}_{3}+\epsilon^{2}_{3}\Omega^{2}_{3}}{\Omega^{2}_{1}+\Omega^{2}_{2}+\Omega^{2}_{3}}} \nonumber \\
    &=& \left[1 + \dfrac{2\epsilon_{2}\Omega^{2}_{2}+2\epsilon_{3}\Omega^{2}_{3}}{\Omega^{2}_{1}+\Omega^{2}_{2}+\Omega^{2}_{3}} + \dfrac{\epsilon^{2}_{2}\Omega^{2}_{2}+\epsilon^{2}_{3}\Omega^{2}_{3}}{\Omega^{2}_{1}+\Omega^{2}_{2}+\Omega^{2}_{3}} \right]^{1/2}.
\label{eq:L_Omega_I_1}
\end{eqnarray}
The Taylor expansion of equation (\ref{eq:L_Omega_I_1}) gives
\begin{eqnarray}
    \dfrac{L}{\Omega I_{1}} &=& 1 + \dfrac{1}{2} \cdot \left( \dfrac{2\epsilon_{2}\Omega^{2}_{2}+2\epsilon_{3}\Omega^{2}_{3}}{\Omega^{2}_{1}+\Omega^{2}_{2}+\Omega^{2}_{3}} + \dfrac{\epsilon^{2}_{2}\Omega^{2}_{2}+\epsilon^{2}_{3}\Omega^{2}_{3}}{\Omega^{2}_{1}+\Omega^{2}_{2}+\Omega^{2}_{3}} \right) - \dfrac{1}{8} \cdot \left( \dfrac{2\epsilon_{2}\Omega^{2}_{2}+2\epsilon_{3}\Omega^{2}_{3}}{\Omega^{2}_{1}+\Omega^{2}_{2}+\Omega^{2}_{3}} + \dfrac{\epsilon^{2}_{2}\Omega^{2}_{2}+\epsilon^{2}_{3}\Omega^{2}_{3}}{\Omega^{2}_{1}+\Omega^{2}_{2}+\Omega^{2}_{3}} \right)^{2} + \cdots \nonumber \\
    &=& 1 + \dfrac{1}{2} \cdot (2\epsilon_{2}\hat{\Omega}^2_{2}+2\epsilon_{3}\hat{\Omega}^2_{3}+\epsilon^{2}_{2}\hat{\Omega}^{2}_{2}+\epsilon^{2}_{3}\hat{\Omega}^{2}_{3}) - \dfrac{1}{8} \cdot (2\epsilon_{2}\hat{\Omega}^2_{2}+2\epsilon_{3}\hat{\Omega}^2_{3}+\epsilon^{2}_{2}\hat{\Omega}^{2}_{2}+\epsilon^{2}_{3}\hat{\Omega}^{2}_{3})^{2} + \cdots
\label{eq:L_Omega_I_1_2}
\end{eqnarray}
The Taylor expansion of the second term on the right-hand side of equation (\ref{eq:cos_hat_theta}) gives 
\begin{eqnarray}
    \hat{L}^{2}_{1} + \dfrac{\hat{L}^{2}_{2}}{1+\epsilon_{2}} + \dfrac{\hat{L}^{2}_{3}}{1+\epsilon_{3}} = \hat{L}^{2}_{1} + \hat{L}^{2}_{2}\cdot(1-\epsilon_{2}+\epsilon^{2}_{2}+\cdots) + \hat{L}^{2}_{3}\cdot(1-\epsilon_{3}+\epsilon^{2}_{3}+\cdots)
\label{eq:hat_L1_L2_L3}
\end{eqnarray}
Substituting equation (\ref{eq:L_Omega_I_1_2}) and equation (\ref{eq:hat_L1_L2_L3}) into equation (\ref{eq:cos_hat_theta}), one can obtain that
\begin{eqnarray}
  \cos{\hat{\theta}} = \boldsymbol{\hat{L}}\cdot\boldsymbol{\hat{\Omega}} &= \left[ 1 + \dfrac{1}{2} \cdot (2\epsilon_{2}\hat{\Omega}^2_{2}+2\epsilon_{3}\hat{\Omega}^2_{3}+\epsilon^{2}_{2}\hat{\Omega}^{2}_{2}+\epsilon^{2}_{3}\hat{\Omega}^{2}_{3}) - \dfrac{1}{8} \cdot (2\epsilon_{2}\hat{\Omega}^2_{2}+2\epsilon_{3}\hat{\Omega}^2_{3}+\epsilon^{2}_{2}\hat{\Omega}^{2}_{2}+\epsilon^{2}_{3}\hat{\Omega}^{2}_{3})^{2} + \cdots \right] \nonumber \\ 
  &\quad\cdot \left[ \hat{L}^{2}_{1} + \hat{L}^{2}_{2}\cdot(1-\epsilon_{2}+\epsilon^{2}_{2}+\cdots) + \hat{L}^{2}_{3}\cdot(1-\epsilon_{3}+\epsilon^{2}_{3}+\cdots) \right].
\label{eq:cos_hat_theta_2}  
\end{eqnarray}
Since we are studying the case where $\epsilon_{2} \ll 1$ and $\epsilon_{3} \ll 1$, we neglect the second-order and higher-order terms in $\epsilon_{2}$ and $\epsilon_{3}$, as well as terms containing $\epsilon_{2} \cdot \epsilon_{3}$, retaining only the first-order terms in $\epsilon_2$ and $\epsilon_3$. Thus, equation (\ref{eq:cos_hat_theta_2}) can be transformed into
\begin{equation}
    \cos{\hat{\theta}} = \boldsymbol{\hat{L}}\cdot\boldsymbol{\hat{\Omega}} \simeq 1 - \epsilon_{2}\hat{L}^{2}_{2} - \epsilon_{3}\hat{L}^{2}_{3} + \epsilon_{2}\hat{\Omega}^{2}_{2} + \epsilon_{3}\hat{\Omega}^{2}_{3}\,,
\label{eq:hat_cos_theta_3}
\end{equation}
which means the angle $\hat{\theta}$ is very close to zero.

In our work, the minimum and maximum values obtained for $\epsilon_2$ are $1.538 \times10^{-6}$ and $21.539\times10^{-6}$, respectively. For $\epsilon_3 = \epsilon_2 + \xi$, the minimum and maximum values are $4.432\times10^{-6}$ and $47.357 \times10^{-6}$, respectively. Therefore, it is reasonable to adopt the approximation that the angular velocity vector is aligned with the angular momentum vector when calculating the cosine of the magnetic inclination angle $\alpha$, i.e., in the calculation of equation (\ref{eq:cos_alpha}).

\end{document}